\documentclass[aps,pra,twocolumn,groupedaddress]{revtex4-2}
\usepackage{blindtext}
\usepackage{enumitem}
\usepackage{braket}
\usepackage{float}
\usepackage{amssymb}
\usepackage{stackengine}
\usepackage{mathrsfs}
\usepackage{ stmaryrd }
\usepackage{ bbold }
\usepackage{graphicx}
\usepackage{epstopdf}
\usepackage{amsmath}
\usepackage[dvipsnames]{xcolor}
\usepackage{tikz}
\usepackage[bottom=0.95in, right=0.62in]{geometry}

\newcommand{\ssymbol}[1]{^{\@fnsymbol{#1}}}
\input{alphabet.tex}
\usepackage{ dsfont }
\begin{document}

% Use the \preprint command to place your local institutional report
% number in the upper righthand corner of the title page in preprint mode.
% Multiple \preprint commands are allowed.
% Use the 'preprintnumbers' class option to override journal defaults
% to display numbers if necessary
%\preprint{}

%Title of paper
\title{Pure State Tomography with Parallel Unentangled  Measurements}

% repeat the \author .. \affiliation  etc. as needed
% \email, \thanks, \homepage, \altaffiliation all apply to the current
% author. Explanatory text should go in the []'s, actual e-mail
% address or url should go in the {}'s for \email and \homepage.
% Please use the appropriate macro foreach each type of information

% \affiliation command applies to all authors since the last
% \affiliation command. The \affiliation command should follow the
% other information
% \affiliation can be followed by \email, \homepage, \thanks as well.
\author{François Verdeil}
%\email[]{Your e-mail address}
%\homepage[]{Your web page}
%\thanks{}
%\altaffiliation{}
\affiliation{Université de Toulouse, UPS, CNRS, CNES, OMP, IRAP, Toulouse, France}
\author{Yannick Deville}
%\email[]{Your e-mail address}
%\homepage[]{Your web page}
%\thanks{}
%\altaffiliation{}
\affiliation{Université de Toulouse, UPS, CNRS, CNES, OMP, IRAP, Toulouse, France}
%Collaboration name if desired (requires use of superscriptaddress
%option in \documentclass). \noaffiliation is required (may also be
%used with the \author command).
%\collaboration can be followed by \email, \homepage, \thanks as well.
%\collaboration{}
%\noaffiliation

\date{\today}

\begin{abstract}
Quantum state tomography (QST) aims at estimating a quantum state from averaged quantum measurements made on copies of the state. Most quantum algorithms rely on QST at some point and it is a well explored topic in the literature, mostly for mixed states. In this paper we focus on the QST of a pure quantum state using parallel unentangled measurements. Pure states are a small but useful subset of all quantum states, their tomography requires fewer measurements and is essentially a phase recovery problem. Parallel unentangled  measurements are easy to implement in practice because they allow the user to measure each qubit individually. We propose two sets of quantum measurements that one can make on a pure state as well as the algorithms that use the measurements outcomes in order to identify the state. We also discuss how those estimates can be fined tuned by finding the state that maximizes the likelihood of the measurements with different variants of the likelihood. The performances of the proposed three types of QST methods are validated by means of detailed numerical tests.
%We will also give an application of our algorithm for quantum process tomography (QPT). QPT aims at identifying a given quantum process, it is a major quantum information processing tool, since it especially allows one to characterize the actual behavior of quantum gates, which are the building blocks of quantum computers. 
                 \end{abstract}
%\maketitle must follow title, authors, abstract, and keywords
\newcommand\barbelow[1]{\stackunder[1.2pt]{$#1$}{\rule{1.3ex}{.13ex}}}
\maketitle
\section{Prior Work and Problem Statement}
Quantum state tomography (QST) aims at estimating a quantum state from averaged quantum measurements made on copies of the state. It often is a necessary step in quantum computation \cite{bookNC}, it has been extensively studied for mixed states. The most basic version is detailed in \cite{bookNC} at the beginning of Section 8.4.2, it uses measurements defined by Pauli operators, often called Pauli measurements (\cite{natureCS}, \cite{CS2}, \cite{CS3}, \cite{PRA1}, \cite{Paulisuite1} and \cite{Paulisuite2}). This version is simple and very robust but requires computing the averages of $4^{n_{qb}}-1$ different types of 2-outcome measurements where $n_{qb}$ is the number of qubits of the state. This scales really badly with the number of qubits but requiring so many types of measurements is not surprising because an arbitrary state is represented by a $d \times d$ Hermitian density matrix with $4^{n_{qb}}$ real parameters (where $d=2^{n_{qb}}$ is the dimension of the Hilbert space in which the considered state evolves). In order to perform QST with fewer types of measurements, one can focus on a subset of all states. The most popular assumption is that the density matrix $\rho$ representing the state has a low rank. \cite{CS3} introduced a compressed sensing approach that requires the averages of $O(r d \log(d)^2)$ 2-outcome measurements to estimate the state where $r$ is the rank of $\rho$. \cite{CS2}, \cite{natureCS}, later built upon this idea of QST via compressed sensing. More recently bounded rank QST was introduced \cite{Brank}. It assumes that the rank $r$ is known and allows the explicit reconstruction of $\rho$ using predetermined measurements (contrary to the compressed sensing approach of \cite{CS3} that does not specify the measurements to be used and finds $\rho$ by minimizing the nuclear norm of $\rho$ under constraints). 

Other approaches do not make any assumption on $\rho$. In 2014 Self-Guided Quantum Tomography (SGQT) was introduced \cite{SGQT} and further studied in \cite{SGQT2}, \cite{SGQT3}. It makes no assumption on $\rho$, and the number of measurements scales reasonably with the number of qubits. The drawback of SGQT is that the measurements that need to be performed on the state are not known beforehand and are generally entangled measurements. Entangled measurements correspond to multiqubit operators that cannot be expressed as a tensor product of single-qubit measurement operators i.e. they cannot be performed by measuring each qubit independently. In 2020 \cite{QOT} introduced a method to partially identify large quantum systems (more than 100 qubits) with entangled states, for which the total state cannot even be stored on a classical computer. It relies on unentangled measurements which are easier to perform than entangled measurements in practice.

The present paper focuses on the tomography of pure states using unentangled measurements. This has been studied in \cite{PRA1} which tried to find the minimal number of Pauli measurements for 2 and 3 qubits (Pauli measurements are unentangled). Our addition to that article is that we will address the generic case with any $n_{qb}$. Furthermore, we will use parallel measurements like in \cite{QOT} where it is shown that all $4^{n_{qb}}$ averaged Pauli measurements can be computed from the averages of $3^{n_{qb}}$ parallel unentangled measurements. A parallel measurement has $d$ outcomes and provides more information on the system than a Pauli measurement that only has two outcomes.

In \cite{Finkelstein} Finkelstein describes a setup able to distinguish almost all pure states, with only $n_{prob}=2d$ probabilities. \cite{Finkelstein} does not beat the lower bound of \cite{finlandais}, detailled below, because there is a negligible (zero measure) set of pure states that the setup of \cite{Finkelstein} cannot recover up to a global phase, it is called the failure set. In addition to the failure set, the main problem of \cite{Finkelstein} is that the measurements are not practical, they are entangled and cannot be performed in parallel (as the matrix $\Av$ associated with the measurements cannot be written as the vertical concatenation of unitary matrices). In \cite{Goyeneche} Goyeneche et al. introduced a set of $n_{prob}=4d$ probabilities that also has a negligible failure set. Technically \cite{Goyeneche} introduces $5$ measurements that yield $5d$ probabilities (obtained from averaging the results of $5$ different kinds of $d$-outcome measurements), but only the $4$ measurements defined by its Equation (2) are needed to achieve QST. The measurements of \cite{Goyeneche} are more realistic as they are performed on 4 orthonormal bases. Two of them are unentangled but the other two are entangled. Goyeneche et al. acknowledge that this is a problem and point out the fact that the two entangled bases can be mapped into the two unentangled ones by applying the quantum Fourier transform twice. In practice this would introduce additional errors, as there are no error-free circuits able to perform the quantum Fourier transform, and one would need to perform quantum process tomography (which generally relies on QST) in order to quantify the errors and improve the Fourier-transform circuit. This is a common issue with entangled measurements, the easiest way to perform them with the current version of quantum computers is to transform them into measurements in an unentangled basis, by means of a corresponding quantum gate. 

The applied mathematics community also dealt with an equivent version of the QST problem for pure states: The phase retrieval problem (see \cite{finlandais}, \cite{4dmoins2}, \cite{mallat}, \cite{injectivity}, \cite{Finkelstein}). A pure state $\ket{\varphi}$ of an $n_{qb}$-qubit system is represented by a complex unit-norm vector $\vb$ with $d$ elements. Pure state tomography aims at estimating $\vb$ from measurements. The theoretical probabilities of all outcomes of the considered types of measurements are contained in the vector $|\Av \vb|^2$ where $\Av$ is an $ n_{prob}\times d$ matrix ($n_{prob}$ is the total number of probabilities) determined by the types of measurements performed and $|.|^2$ stands for component-wise squared modulus. Recovering $\vb$ (up to a global phase) from $|\Av \vb|^2$ (generally it is $|\Av \vb|$ instead of $|\Av \vb|^2$ but both problems are essentially the same) is called phase retrieval. The first question asked in phase recovery is injectivity: how can one choose $\Av$ in order to make sure that $|\Av \vb|^2$ contains enough information to recover $\vb$ up to a global phase? Proving that a given $\Av$ guarantees injectivity is a difficult question. \cite{finlandais} gave a minimal number of measurements below which injectivity is impossible. In our case this condition is $n_{prob}>4d-3-c(d)n_{qb}$ rows for some $c(d)\in [1, 2]$. \cite{4dmoins2} showed that for a generic $\Av$, having $4d-2$ rows or more is a sufficient condition for injectivity. 
%However, its definition of "generic" is abstract: A must be in an unspecified open and dense subset of all $n_{prob}\times 2^{n_{qb}}$ matrices. 

Beyond injectivity, finding a solution to the phase recovery problem (whether it is unique up to a global phase or not) is the main difficulty of pure state tomography. Both \cite{Finkelstein} and \cite{Goyeneche} give their own closed-form algorithms to recover the phases which are adapted to their versions of $\Av$. \cite{mallat} focuses on this particular problem with a generic $\Av$.

Our contributions in the present paper are as follows.
Section~\ref{section:DataModel} describes the quantum state to be identified and the measurements made. In particular, we formalize the definition of a parallel unentangled measurement.

Section~\ref{section:petitA} describes a method to achieve QST with $n_{prob}=4d$ using an optimization algorithm of \cite{mallat} on a number of probabilities consistent with the lower bound of \cite{finlandais}. The probabilities can be obtained by averaging the results of 4 types of parallel unentangled measurements.

Section~\ref{section:grandA} describes an original method with $n_{prob}=(2n_{qb}+1)d$ probabilities for which phase recovery can be achieved with a closed-form recursive algorithm. Those probabilities are obtained by averaging the results of $2n_{qb}+1$ different kinds of measurements. 

Section~\ref{section:ML} describes a more precise fine tuning method that works with all types of measurements, it requires an initial estimate from one of the algorithms of Sections \ref{section:petitA} or \ref{section:grandA} which it uses in order to maximize the likelihood of the measurements. 

Finally, in Section \ref{section:plot} we evaluate the performance of the proposed algorithms with simulated data.
\section{State and measurements}
\label{section:DataModel}
\subsection{Considered state}
\label{section:State}
An $n_{qb}$-qubit pure state $\ket{\varphi}$ can be decomposed in the canonical basis $\ket{0...0}$, ..., $\ket{1...1}$. The components of $\ket{\varphi}$ in the basis can be stored in a $d$-element vector ($d = 2^{n_{qb}}$)
$
\vb = \begin{bmatrix}
v_1 & ... & v_d
\end{bmatrix}^T
$
where $^T$ stands for transpose. The components $v_j$ are complex and $\sum_{j=1}^{d}{|v_j|^2}=1$. The global phase of $\ket{\varphi}$ has no physical meaning, so we can assume that $v_1$ is a real non-negative number.
\subsection{Projective measurement}
\label{section:mesGen}
According to Section 2.2.5 in \cite{bookNC} a projective measurement is defined by a Hermitian matrix \Bv whose distinct eigenvalues $m_k$ are the possible outcomes of the measurement. The probability of getting $m_k$ when measuring a pure state represented by $\vb$ is $\varrho_k(\vb)$: the squared norm of the projection of $\vb$ on the eigenspace associated with $m_k$. 

A projective measurement can have the following properties:

\begin{itemize}[leftmargin=3.5mm]
\item $d$-outcome: If \Bv has $d$ distincts eigenvalues. The eigenstates are then $1$-dimensional spaces, and $\varrho_k(\vb)=|\eb_k^* \vb|^2$ where $\eb_k$ is a $d$-element unit-norm vector belonging to the $k$-th eigenspace of $\Bv$ ($^*$ is the transconjugate). By performing the measurement several times on copies of the state, we can therefore estimate $\{|\eb_k^* \vb|^2\}_{1 \leq k \leq d}$ where $\eb_k$ spans an orthonormal basis. 
%\noindent This type of measurement is strictly better than one that yields fewer than $d$ outcomes in the following sense: For each $j$-outcome measurement $\mathcal{M}_j$ with $j<d$ there exit a $d$-outcome measurement $\mathcal{M}_d$ such that any two states $\sbv{v_1}$ and $\sbv{v_2}$ that can be distinguished by  $\mathcal{M}_j$ (in the sense that the probabilities of each outcome are not all the same for $\sbv{v_1}$ and $\sbv{v_2}$) can also be distinguished by  $\mathcal{M}_d$; and the reverse is not true: There exists a pair of states that can be distinguished by $\mathcal{M}_d$ but not by $\mathcal{M}_j$. 
\item Unentangled: If the measurement can be performed with simultaneous local measurements on each qubit. For example if the matrix \Bv can be written as a tensor product of $n_{qb}$ matrices with dimension $2\times 2$: $\Bv = \Bv_1 \otimes ... \otimes \Bv_{n_{qb}}$. Then the measurement represented by \Bv can be performed by measuring simultaneously each qubit with the measurements represented by $\Bv_1, ..., \Bv_{n_{qb}}$ and then computing the product of all the outcomes. Depending on how the 2 eigenvalues of each $\Bv_k$ are chosen, computing the product of the outcomes can result in a loss of information (i.e. the outcomes of each $\Bv_k$ cannot be retrieved from their product knowing their respective 2 possible values). If this is the case \Bv will have fewer than $d$ distinct eigenvalues (see e.g. Pauli measurements).
\end{itemize}
\subsection{Parallel unentangled measurement}
\label{section:mes}
If a quantum measurement represented by $\Bv$ is unentangled and has $d$ outcomes, we call it a parallel unentangled measurement. It can be performed on single qubits in parallel using the following setup:

\tikzset{every picture/.style={line width=0.75pt}} %set default line width to 0.75pt        

\begin{tikzpicture}[x=0.75pt,y=0.75pt,yscale=-1,xscale=1]
%uncomment if require: \path (0,392); %set diagram left start at 0, and has height of 392

%Straight Lines [id:da9116191926973902] 
\draw    (38,138.33) -- (75,138.64) ;
\draw [shift={(75,138.67)}, rotate = 180.64] [color={rgb, 255:red, 0; green, 0; blue, 0 }  ][line width=0.75]    (10.93,-3.29) .. controls (6.95,-1.4) and (3.31,-0.3) .. (0,0) .. controls (3.31,0.3) and (6.95,1.4) .. (10.93,3.29)   ;
%Straight Lines [id:da34636628780590395] 
\draw    (38,166.33+10) -- (75,166.64+10) ;
\draw [shift={(75,166.67+10)}, rotate = 180.64] [color={rgb, 255:red, 0; green, 0; blue, 0 }  ][line width=0.75]    (10.93,-3.29) .. controls (6.95,-1.4) and (3.31,-0.3) .. (0,0) .. controls (3.31,0.3) and (6.95,1.4) .. (10.93,3.29)   ;
%Straight Lines [id:da253003226382272] 
\draw    (38,215.33+10) -- (75,215.64+10) ;
\draw [shift={(75,215.67+10)}, rotate = 180.64] [color={rgb, 255:red, 0; green, 0; blue, 0 }  ][line width=0.75]    (10.93,-3.29) .. controls (6.95,-1.4) and (3.31,-0.3) .. (0,0) .. controls (3.31,0.3) and (6.95,1.4) .. (10.93,3.29)   ;

% Text Node
\draw (18,175+11) node [anchor=north west][inner sep=0.75pt]   [align=left] {\vdots};

% Text Node
\draw (43,132-12) node [anchor=north west][inner sep=0.75pt]    {$\mathcal{M}_1$};
% Text Node
\draw (43,159-12+10) node [anchor=north west][inner sep=0.75pt]    {$\mathcal{M}_2$};
% Text Node
\draw (43,210-15+10) node [anchor=north west][inner sep=0.75pt]    {$\mathcal{M}_{n_{qb}}$};

% Text Node
\draw (10,129.4+3) node [anchor=north west][inner sep=0.75pt]    {$q_1$};
% Text Node
\draw (10,157.4+10+3) node [anchor=north west][inner sep=0.75pt]    {$q_2$};
% Text Node
\draw (10,207.4+10+3) node [anchor=north west][inner sep=0.75pt]    {$q_{n_{qb}}$};

% Text Node
\draw (85,129.4+3) node [anchor=north west][inner sep=0.75pt]    {2 outcomes};%{$\{ m_1^1, m_1^2\}$};
% Text Node
\draw (85,157.4+10+3) node [anchor=north west][inner sep=0.75pt]    {2 outcomes};%{$\{ m_2^1, m_2^2\}$};

\draw (22+75+13,175+11) node [anchor=north west][inner sep=0.75pt]   [align=left] {\vdots};
% Text Node
\draw (85,207.4+10+3) node [anchor=north west][inner sep=0.75pt]    {2 outcomes};%{$\{ m_d^1, m_d^2\}$};

%Curve Lines [id:da03656413654736057] 
\draw [line width=1.5]    (-335+380,222.12+35) .. controls (-335+381.09,220.81+35) and (-335+382.15,219.65+35) .. (-335+383.19,218.64+35) .. controls (-335+394.26,207.78+35) and (-335+402.46,212.66+35) .. (-335+409.65,222.12+35) ;
%Straight Lines [id:da35240484424797947] 
\draw    (-335+392.6,221.44+35) -- (-335+403.53,206.68+35) ;
\draw   (-335+398.61,208.93+35) -- (-335+403.68,206.44+35) -- (-335+402.27,212.9+35) ;

%Curve Lines [id:da07003512385312116] 
\draw    (259.53-95,282.75-42) .. controls (276.56-95,276.4-42) and (263.42-95,231.24-42) .. (276.38-95,229.93-42) ;
%Curve Lines [id:da5860663743021556] 
\draw    (276.38-95,229.93-42) .. controls (264.86-95,230.61-42) and (276.19-95,180.95-42) .. (259.59-95,172.22-42) ;

\draw (185,180) node [anchor=north west][inner sep=0.75pt]    {$d$ outcomes};

\end{tikzpicture}

\noindent Each qubit $q_1,..., q_{n_{qb}}$ composing the system is measured with a one-qubit measurement which has two distinct outcomes. %and $m_k^1$ and $m_k^2$ are the two eigenvalues of $\sbv{H_{k}}$.

The eigenvalues of $\Bv$ are not important in the present paper. Changing them without changing the eigenvectors would change $\Bv$ but (as long as the eigenvalues remain distinct) the resulting measurement would be equivalent in the sense that it would give the same information; the possible measured values would then depend on $\Bv$ but the associated probabilities would be the same up to a permutation, and this is what matters in the present paper.

Since the parallel unentangled measurements we consider have $d$ outcomes (by definition), $\varrho_k(\vb)=|\eb_k^* \vb|^2$ where $\eb_k$ is a $d$-element unit-norm vector belonging to the $k$-th eigenspace.

For a given parallel unentangled measurement $\mathcal{M}$ let us define the eigenvector matrix: $\Ev_{\mathcal{M}} = \begin{bmatrix}\eb_1 & \hdots &\eb_d \end{bmatrix}$. The vector $\pb_{\mathcal{M}}(\vb) = |\Ev_{\mathcal{M}}^* \vb|^2$ contains the $d$ probabilities of the possible outcomes.
%Measuring a set of $n_{qb}$ qubits in parallel one time gives a string of $n_{qb}$ classical bits which takes one of $d$ values. We define the theoretical probability vector of the measurement $\mathcal{M}$ as a $d$-element vector where the $i$-th string is the probability of observing the i-th sequence of bits (so $(0...0)$ for the first element $(0...01)$ for the second etc...) we call it $\pb_{\mathcal{M}}(\vb)$.
%For each type of measurement $\mathcal{M}$ we can find a unitary matrix $\Ev$ such that:
%$\pb_{\mathcal{M}}(\vb) = |P \vb|^2$ (equivalent to (2.103) in \cite{bookNC} when all the outcomes are distincs which is the case here). We call $\Ev$ the eigenvector matrix of the measurement (it contains the eigenvectors of the Hermitian operator defined in (2.102) of \cite{bookNC}).

By performing several measurements on copies of the state represented by $\vb$, we compute the frequencies of occurrence of each outcome, we get $\widehat{\pb_{\mathcal{M}}}$, which we use as an approximation of $|\Ev_{\mathcal{M}}^* \vb|^2$. We call  $\widehat{\pb_{\mathcal{M}}}$ the averaged measurements or sample probabilities.
The sum of the elements of $\pb_{\mathcal{M}}(\vb)$ is one (it is the sum of the probabilities of all possible outcomes), so no information is lost by removing one element. 
%Since the norm of $\vb$ is 1, and $\Ev$ is unitary, the sum of the element of $|P \vb|^2$ is 1 and even though the theoretical probability of each outcome yields $d$ real numbers, it is only characterized by the $d-1$ first.
We define $\barbelow{\Ev_{\mathcal{M}}}$ the non-redundant eigenvector matrix as composed of the first $d-1$ columns of $\Ev_{\mathcal{M}}$. Then $\pb_{\mathcal{M}}(\vb)=|\Ev_{\mathcal{M}}^* \vb|^2$ is redundant but $\barbelow{\pb_{\mathcal{M}}}(\vb)=|\barbelow{\Ev_{\mathcal{M}}^*} \vb|^2$ is not.  

$\barbelow{\Ev_{\mathcal{M}}^*}$,  $\Ev_{\mathcal{M}}^*$, $\barbelow{\pb_{\mathcal{M}}}(\vb)$ and $\pb_{\mathcal{M}}(\vb)$ will all be used at different points of this article with $\mathcal{M}$ replaced by the actual measurements we will perform.%In Sections \ref{section:petitA}, we are going to use $\barbelow{\Ev_{\mathcal{M}}^*}$ when describing the measurements, this way no redundant information is added. But when we use the sample probabilities to perform the QST, we will use $\widehat{\pb_{\mathcal{M}}}$ and compare it to $|\Ev_{\mathcal{M}}^* \vb|^2$ even though they are redundant.%s, this implies that $|P \vb|^2$ can be deduced from $|\barbelow{P} \vb|^2$.
\subsection{Considered types of measurements}
\label{section:Pauli}
%A Pauli measurement describes the spin of electrons interacting with each other in each of the three spatial directions.
We perform measurements for all qubits in parallel, with one measurement direction per qubit.
For one qubit, we choose to perform measurements that are equivalent (up to a factor 1/2 on the outcomes) to the 3 non-trivial Pauli measurements. The measurement matrices $\Bv$ associated with the directions X, Y and Z are the last three Pauli matrices defined in Section 2.1.3 of \cite{bookNC} and the corresponding eigenvector matrices may be shown to read:
\begin{equation}
\label{eqn:Pauli2}
\Ev_X = \frac{1}{\sqrt{2}} \begin{pmatrix} 1 & 1\\
1 & -1
\end{pmatrix} 
\Ev_Y = \frac{1}{\sqrt{2}} \begin{pmatrix} 1 & 1\\
i & -i
\end{pmatrix} 
\Ev_Z = \begin{pmatrix} 1 & 0\\
0 & 1
\end{pmatrix}.
\end{equation}
If the qubit represents the spin of an electron, those eigenvector matrices represent the measurement of the spin component along 3 orthogonal directions. There is a factor $1/2$ between the outcome of the spin measurements and the Pauli measurements but it does not affect the eigenvectors.

For two or more qubits, the different qubits can be measured along X, Y or Z. It can be shown that the resulting eigenvector matrix is the tensor product of the 2-dimensional matrices of (\ref{eqn:Pauli2}). For example for 2 qubits, measuring the first one along Z and the second one along X has the following eigenvector matrix

$\Ev_{ZX} = \Ev_{Z} \otimes \Ev_{X} =  \frac{1}{\sqrt{2}} \begin{pmatrix} 1 & 1 & 0 & 0\\
1 & -1 & 0 & 0\\
0 & 0 & 1 & 1\\
0 & 0 & 1 & -1
\end{pmatrix}$.

In this example, if the qubits represent the spins of 2 electrons, then the measurement we perform is equivalent  to measuring the first spin component along $Z$ and the second along $X$. The spin measurement has 4 possible outcomes $(+\frac{1}{2}, +\frac{1}{2}), (+\frac{1}{2}, -\frac{1}{2}), (-\frac{1}{2}, +\frac{1}{2})$ and $(-\frac{1}{2}, -\frac{1}{2})$ and if $\vb$ represents the considered state, the probabilities of each outcome are in the vector $|\Ev_{ZX}^* \vb|^2$. This measurement is not equivalent to a two-qubit Pauli measurement (even if we forget the factor $1/2$), as such a Pauli measurement only has two outcomes. In fact, the Pauli measurement along $ZX$ would return $+1$ for spins measurement outcomes $(+\frac{1}{2}, +\frac{1}{2})$ and $(-\frac{1}{2}, -\frac{1}{2})$ and $-1$ for spins measurement outcomes $(+\frac{1}{2}, -\frac{1}{2})$ and $(-\frac{1}{2}, +\frac{1}{2})$. This is inefficient as half the information is wasted.

For $n_{qb}$ qubits there are $3^{n_{qb}}$ different measurements of this type. Both of the QST methods of Section \ref{section:petitA} and \ref{section:grandA} as well as the fine tuning algorithms of Section \ref{section:ML} use a specific subset of all possible measurements. 
\subsection{Justification}
\label{section:Pauli2}

We think that performing QST using a kind of measurement that is not parallel unentangled (i.e. has fewer than $d$ outcomes or is entangled) should not be recommended in practice with the current state of quantum computers for the following reasons:
\begin{itemize}[leftmargin=3.5mm]
\item Performing a quantum measurement that has fewer than $d$ outcomes is suboptimal. Indeed, instead of considering a $j$-outcome measurement $\mathcal{M}_j$ ($j<d$) we can use a $d$-outcome measurement $\mathcal{M}_d$ that has the same eigenvectors and $d$ distinct eigenvalues. With this definition it is strictly better to use $\mathcal{M}_d$ than $\mathcal{M}_j$ in all situations, as the outcomes of $\mathcal{M}_d$ can be mapped injectively onto the outcomes of $\mathcal{M}_j$ but the reverse is not true. Therefore $\mathcal{M}_d$ brings us strictly more information on the system than $\mathcal{M}_j$ and performing either of them should be as difficult (a copy of the state is used up).
\item Performing an entangled measurement requires the use of a quantum gate. This gate itself is never going to act exactly as expected and will introduce errors. In order to see if the gate works as expected, we would need to perform quantum process tomography which generally relies on QST.
\end{itemize}

But the literature on QST is full of theoretical papers that consider measurements that fall within the two types that we do not recommend. Here are some examples: 
\begin{itemize}[leftmargin=3.5mm]
\item \cite{SGQT} uses successive 2-outcome projective measurements on non-orthogonal entangled eigenstates. And each iteration of the algorithm would require a new type of measurement (that depends on what has been measured before and is most likely going to be entangled) and therefore a new quantum gate has to be built on the fly.
\item \cite{Finkelstein} considers projective 2-outcome measurements on 1-dimentional spaces. Half of those measurements can be performed using a single parallel unentangled measurement (with the identity matrix as eigenvector matrix); but the other half cannot.
\item \cite{Goyeneche} considers 2 parallel unentangled measurements (called local measurements in \cite{Goyeneche}) and 2 $d$-outcome entangled measurements that can be mapped on the other two using a gate that performs the Fourier transform. This setup is way more reasonable than the others as it requires a single known standard gate.
\item  \cite{natureCS}, \cite{CS2}, \cite{CS3}, \cite{PRA1}, \cite{Paulisuite1} and \cite{Paulisuite2} all use multiqubit Pauli measurements. Multiqubit Pauli measurements have the advantage of being unentangled and also simplify the calculation for the QST of mixed states (see the beginning of Section 8.4.2 in \cite{bookNC}, (8.149) only works for orthogonal sets of matrices with respect to the Hilbert–Schmidt inner product, like Pauli matrices). They have the disadvantage of being 2-outcome measurements returning either +1 or -1. There are sets of Pauli measurements whose expected values can be deduced from the outcomes of parallel unentangled measurements without loss of information (\cite{QOT} explains how it can be done for two qubits). But that is not the case for any set of Pauli measurements.
\end{itemize}
%The easiest way to generate all measured values of the $4^{n_{qb}}-1$ non-trivial Pauli measurements in practice is to perform all the $3^{n_{qb}}$ combinations of 1-qubit $X, Y$ and $Z$ Pauli measurements on each qubit (i.e. the measurements described above), and then compute the products of the right 1-qubit measured values (\cite{QOT} details a related approach for $2$ qubits). It can be useful to deal with this complication because Pauli measurements are so simple for mixed states, but as we use pure states, we choose to ignore the issue.

In contrast to those articles we here make a point to only use unentangled parallel measurements. We could have chosen other matrices than (\ref{eqn:Pauli2}). We chose those matrices in order to be closer to the Pauli measurements widely used in the literature.
\section{Tomography with minimal number of measurement types}
\label{section:petitA}
The current section describes our first QST setup, Section \ref{section:mess} describes the 4 types of parallel unentangled measurements that are performed, Section \ref{section:injectivity} explains why it is reasonable to think that they are injective up to a global phase and Section \ref{section:PR} describes a first algorithm to recover the phases.
\subsection{Types of measurements}
\label{section:mess}
In the QST method described here, we perform 4 types of measurements on the considered $d$-dimensional state: The first measurement measures all the qubits along Z, its eigenvector matrix, $\Ev_{Z...Z}$ is the identity matrix, the second measurement measures all the qubits along Y, the third along X, and the fourth measures every odd-numbered qubit along X and every even-numbered qubit along Y.

After performing the measurements several times on copies of the state, we compute the sample probabilities $\widehat{\pb_{\mathcal{M}}}$ for $\mathcal{M}$ spanning the 4 types of measurements. We then have an $n_{s} = 4d$ dimensional vector with $\barbelow{n_{s}} = 4(d-1)$ degrees of freedom. We call it $\widehat{\pb_s}$. The associated theoretical probability vector is $\pb_s=|\Av_{s} \vb|^2$, where $s$ stands for ``small" because the corresponding matrix in Section \ref{section:grandA} has more rows. $\Av_{s}$ is the concatenation of the transconjugates of the eigenvector matrices of the measurements we perform, $\barbelow{\Av_{s}}$ is defined similarly
\begin{equation}
\label{eqn:Asmall}
\Av_{s} = \begin{bmatrix} \Ev_{Z...Z}^* \\\Ev_{Y...Y}^* \\ \Ev_{X...X}^* \\ \Ev_{XYXY...}^* \end{bmatrix}  \text{ and }  \barbelow{\Av_{s}} = \begin{bmatrix} \barbelow{\Ev_{Z...Z}^*} \\ \barbelow{\Ev_{Y...Y}^*} \\ \barbelow{\Ev_{X...X}^*} \\ \barbelow{\Ev_{XYXY...}^*} \end{bmatrix}.
\end{equation}
Let us define $\barbelow{\pb_s}=|\barbelow{\Av_{s}} \vb|^2$. Since the norm of $\vb$ is 1, $\barbelow{\pb_s}$ and $\pb_s$ contain the same information (see Section \ref{section:mes}). In Section \ref{section:injectivity} we will consider $\barbelow{\Av_{s}}$, $\barbelow{n_{s}}$ and $\barbelow{\pb_s}$ in order to see if the measurements are injective because we do not want to introduce redundancy when counting the measurements. But, for the sake of simplicity, we will consider $\Av_{s}$, $n_{s}$ and $\pb_s$ in Section \ref{section:PR} in order to recover the state from the measurements. We want to use all the measurements from $\widehat{\pb_s}$ whether they are redundant or not.
\subsection{Injectivity}
\label{section:injectivity}
$\barbelow{\Av_{s}}$ is an $\barbelow{n_{s}} \times d$ matrix and $\vb$ has unit norm. We want to know whether the measurements we chose are sufficient to recover any $\vb$ from $|\barbelow{\Av_{s}} \vb|^2$ up to a global phase. In the rest of the paper this property will be called injectivity. It is a bit of an exaggeration because $ \vb \rightarrow |\barbelow{\Av_{s}} \vb|^2$ is never truly injective as changing the global phase of \vb will not change $|\barbelow{\Av_{s}} \vb|^2$. This issue of injectivity was studied before in \cite{finlandais}, \cite{4dmoins2}, \cite{injectivity} in a slightly different setup: the considered measurements are $|\barbelow{\Av_{s}} \vb|$ instead of $|\barbelow{\Av_{s}} \vb|^2$ , this does not change anything for the injectivity, also $\vb$ is not assumed to have unit norm, and this is important. In order to reconcile the two setups we can relax the unit-norm hypothesis for $\vb$ and insert the row $[0, ..., 0, 1]$ between the $(d-1)$-th row and the $d$-th row of $\barbelow{\Av_{s}}$. This ensures that the norm of $\vb$ is constrained: its square is the sum of the first $d$ constrained measurements, because the first $d$ rows of $\barbelow{\Av_{s}}$ are the identity matrix. With this change $\barbelow{\Av_{s}}$ has $4d-3$ rows.

According to \cite{finlandais} the minimal number of rows for $\barbelow{\Av_{s}}$ below which injectivity is impossible is $4d-3-c(d)n_{qb}$ rows for some $c(d)\in [1, 2]$. Since we have $4d-3$ rows, this necessary condition is satisfied. However there is no simple sufficient condition on $\barbelow{\Av_{s}}$ that ensures injectivity, and proving it for a given $\barbelow{\Av_{s}}$ is a known hard problem. The closest result we found to a sufficient condition is in \cite{4dmoins2} where it is shown that for a generic $\barbelow{\Av_{s}}$, having $4d-2$ or more rows ensures injectivity. $\barbelow{\Av_{s}}$ must be generic in the sense that it is part of a specific open dense set with full measure. We cannot identify this set and check that $\barbelow{\Av_{s}}$ would be in it (although it probably would because the set is of full measure), but this is moot because we are one row short of satisfying the $4d-2$ condition anyway. However \cite{injectivity} explained why it is natural to think that $4d-4$ is the actual lower bound. It remains a conjecture though. 

We can be sure that 3 measurement types would not be enough to achieve injectivity with $n_{qb}>2$ as the bound of \cite{finlandais} would not be fulfilled: we would have $3d-2$ independent rows ($3d-3$ plus the unit-norm constraint). This is always strictly smaller than $4d-3-2n_{qb}$ for $n_{qb}>2$. 4 is the lowest number of measurement types for which we can hope to always achieve injectivity.
\subsection{A first quantum pure state tomography method}
\label{section:PR}
In the current section, we show how the method proposed in \cite{mallat} can be used in our framework to recover $\vb$ from the sample probabilities $\widehat{\pb_s}$, an estimate of $\pb_s = |\Av_{s} \vb|^2$ (we only consider $\Av_{s}$ from now on, $\barbelow{\Av_{s}}$ was only useful to discuss the injectivity). The optimization problem considered in \cite{mallat} is the following:
\begin{equation}
\label{eqn:pb_base}
 \underset{\vb}{\min} \left\lVert |\Av_{s} \vb|-\sqrt{\widehat{\pb_s}}\right\rVert%\text{ s.t. }\lVert v\rVert = 1
\end{equation}
where $\sqrt{\widehat{\pb_s}}$ is the element-wise square root of $\widehat{\pb_s}$ and $||.||$ is the $L_2$ norm. \cite{mallat} does not include the unit-norm constraint on \vb but, since we use $\Av_{s}$, this constraint is implicit in the criterion to be minimized. In fact, the sum of the first $d$ elements of $|\Av_{s} \vb|^2$ is the squared norm of $\vb$ and the sum of the first $d$ elements of $\widehat{\pb_s}$ is one, therefore if $|\Av_{s} \vb|$ is close to $\sqrt{\widehat{\pb_s}}$, their squared norms will also be close, and therefore the squared norm of $\vb$ will be close to $1$.
In \cite{mallat}, it is shown that (\ref{eqn:pb_base}) is equivalent to the following optimization problem (originally it came from \cite{shor87}): 

\begin{equation}
\label{eqn:reformulation}
 \underset{\Uv\text{ s.t. } \mathcal{C}}{\min} tr(\Uv \Mv)
\end{equation}
where $\sbv{M} = diag(\widehat{\pb_s})(I-\Av_{s}\Av_{s}^{\dagger})diag(\widehat{\pb_s})$,  $^{\dagger}$ is the pseudo-inverse, $diag(\widehat{\pb_s})$ is the diagonal matrix whose diagonal is $\widehat{\pb_s}$ and $\mathcal{C}$ represents the following condition on the $n_{s}\times n_{s}$ matrix $\Uv$: 
\begin{equation}
\label{eqn:defu}
\exists \ub \in \mathds{C}^{n_s} \text{such that } |\ub| = [1, ..., 1]^T \text{and } \Uv = \ub \ub^* .
\end{equation}

\cite{mallat} shows that if $\Uv$ is a solution of (\ref{eqn:reformulation}), then the associated $\ub$ of (\ref{eqn:defu}) is an approximation of the phase of $\Av_{s} \vb$, and the resulting estimate of $\vb$ defined as:
\begin{equation}
\label{eqn:v_cvx}
\widehat{\vb}_{0} = \Av_{s}^{\dagger}(\ub*\sqrt{\widehat{\pb_s}})
\end{equation}
 ($*$ is the element-wise product) is the solution of (\ref{eqn:pb_base}) proposed in \cite{mallat}.

(\ref{eqn:reformulation}) is almost a convex optimization problem. In fact if $\mathcal{C}$ is reformulated in an equivalent way: $\Uv_{i, i} = 1 \text{ } \forall i \in [1, n_{s}], \Uv \succeq 0, Rank(\Uv)=1$ ($\Uv\succeq0$ means that $\Uv$ is both Hermitian and non-negative definite), according to \cite{mallat} the criterion $tr(\Uv \Mv) $ is convex and the only constraint that makes the problem non-convex in $\mathcal{C}$ is $Rank(\Uv)=1$. By relaxing it we have a convex problem that can be solved without the need for a good initialization:
\begin{equation}
\label{eqn:convexe}
 \underset{\Uv\text{ s.t. } \Uv_{i, i} = 1 \forall i, \Uv \succeq 0}{\min} tr(\Uv \Mv) .
\end{equation}
Once (\ref{eqn:convexe}) is solved using the PhaseCut algorithm of \cite{mallat}, the eigenvectors and eigenvalues of the solution \Uv are computed. In order to get an estimate of $\ub$, \cite{mallat} then computes $\widehat{\ub}$, the eigenvector associated with the largest eigenvalue. From $\widehat{\ub}$, we get the estimate of $\vb$ defined in (\ref{eqn:v_cvx}):
\begin{equation}
\label{eqn:v_pc}
\widehat{\vb}_{pc} = \Av_{s}^{\dagger}(\widehat{\ub}*\sqrt{\widehat{\pb_s}}).
\end{equation}
In \cite{mallat} this method is tested with $\Av$ matrices which represent usual use-cases in the signal/image processing community (oversampled Fourier transform, multiple random illumination filters, wavelet transform) for which PhaseCut works well. However for $\Av = \Av_{s}$, PhaseCut is a good initial point but needs the fine tuning that we will detail in Section \ref{section:ML}.
\subsection{Comparison with the literature}
\label{section:compLit1}

Let us sum up the main features of our first QST algorithm:
\begin{itemize}
\item It uses $4d$ probabilities that can be obtained by averaging the results of $4$ parallel unentangled measurements.
\item It is reasonable to think that the chosen measurements are injective (the failure set is most likely empty).
\item The algorithm that reconstructs the state is not explicit (optimization).
\end{itemize}
Goyeneche el al. \cite{Goyeneche} uses the same number of measurement types, has a known failure space of zero measure and provides an explicit reconstruction algorithm. The main advantage our approach based on PhaseCut has compared to \cite{Goyeneche} is that we do not use unentangled measurements. The more general compressed sensing approach of \cite{CS3} requires $O(r d \log(d)^2)$ probabilities to estimate the state where $r$, the rank of the density matrix, is $1$ in the case of a pure state. Those probabilities could be  obtained by averaging the results of $O(\log(d)^2)$ different unentangled measurements. Our method is more efficient since we use $4 = O(1)$ different unentangled measurements. Both methods have no theoretical guarantee of injectivity or closed-form solution. The validity of the solution can only be shown in simulations.

\section{closed-form state tomography algorithm}
\label{section:grandA}

\subsection{Alternative types of measurements}
\label{section:mess2}
In the alternative  QST method described here, we perform the following measurements% $2n_{qb}+1$ types of measurements on the $n_{qb}$ qubits. Those measurements are 

\noindent \scalebox{1}[1]{$\bigg\{\underbrace{Z...Z}_\textrm{$n_{qb}$ times}, \Big\{ \underbrace{Z...Z}_\textrm{$n_{qb}-i$ times} S  \underbrace{X...X}_\textrm{$i-1$ times}, \begin{matrix}1\leq i \leq n_{qb} \\ S\in\{X, Y\} \end{matrix}\Big\} \bigg\}$}
%$\Big\{\underbrace{Z...Z}_\textrm{$n_{qb}$ times}, \big\{ \underbrace{Z...Z}_\textrm{$n_{qb}-i$ times} S  \underbrace{X...X}_\textrm{$i-1$ times}, 1\leq i \leq n_{qb}, S\in\{X, Y\} \big\} \Big\}$%$(X...X), (YX...X),$ $(ZX...X), (ZYX...X), ...., (Z...ZX), (Z...ZY), (Z...Z)$. 

The number of types of measurements is $2n_{qb}+1$. The resulting $\Av_{t}$ ($t$ stands for ``tall") matrix has $n_r=d(2n_{qb}+1)$ rows:
%\begin{equation}
%\label{eqn:Atall}
%\resizebox{.96\hsize}{!}{$
%\Av_{t} = [\Ev_{Z...Z}^*^T, \Ev_{Z...ZX}^*^T, \Ev_{Z...ZY}^*^T, ..., \Ev_{X...X}^*^T, \sbv{\Ev_{YX...X}^*}^T]^T$
%}
%\end{equation}
\begin{equation}
\label{eqn:Atall}
\Av_{t} = \begin{bmatrix}\Ev_{Z...Z}^*\\ \Ev_{Z...ZX}^*\\ \Ev_{Z...ZY}^*\\  \vdots \\ \Ev_{X...X}^*\\ \Ev_{YX...X}^* \end{bmatrix}.
\end{equation}

Each measurement is performed several times and we compute the sample probabilities $\widehat{\sbm{p_t}}$ which are estimates of the theoretical probabilities $\sbm{p_t} = |\Av_{t}\vb|^2$.

$2n_{qb}+1$ sounds like a lot compared to the 4 measurement types of Section \ref{section:petitA} but it is a small fraction of the $3^{n_{qb}}$ possible types of  measurements defined in Section \ref{section:Pauli}. This setup also has the advantage of coming with an attractive way to recover the state from the measurements, as will be explained in Section \ref{section:pinj}.

\subsection{A recursive pure quantum state tomography method}
\label{section:pinj}

Let us show how a vector $\vb$ can be recovered up to a global phase from $|\Av_{t} \vb|^2$ by induction on the number of qubits.

$\Av_{t}$ depends on $n_{qb}$, in the rest of the current section this dependence will not be omitted and $\Av_{t}$  will be called $\Av_{t}(n_{qb})$. We first show how to solve the problem (recover $\vb$ from $|\Av_{t} \vb|^2$) with $n_{qb}=1$. We then explain how solving the problem for $n_{qb}-1$ qubits yields the solution for $n_{qb}$ qubits. From there a recursive algorithm can be implemented.

$n_{qb} = 1$: $\Av_{t}(1) = \begin{bmatrix} \Ev_Z^* \\ \Ev_X^* \\ \Ev_Y^* \end{bmatrix}$, with the $\Ev_Z, \Ev_X, \Ev_Y$ of (\ref{eqn:Pauli2}). 
The state vector is $\vb = \begin{pmatrix} |v_1|\\ |v_2|e^{i \theta} \end{pmatrix}$. 
Basic calculations show: 
\begin{equation}
\label{eqn:A1}
|\Av_{t}(1) \vb|^2=\begin{pmatrix} |v_1|^2 \\ |v_2|^2 \\
\frac{1}{2}\left( |v_1|^2+|v_2|^2+2|v_1||v_2|\cos(\theta)\right)  \\ 
\frac{1}{2}\left( |v_1|^2+|v_2|^2-2|v_1||v_2|\cos(\theta)\right)   \\
\frac{1}{2}\left( |v_1|^2+|v_2|^2+2|v_1||v_2|\sin(\theta)\right)   \\ 
\frac{1}{2}\left( |v_1|^2+|v_2|^2-2|v_1||v_2|\sin(\theta)\right)
\end{pmatrix}.
\end{equation}
Therefore, $|\Av_{t}(1) \vb|^2$ gives $|v_1|^2$, $|v_2|^2$, $|v_1||v_2|\cos(\theta)$ and $|v_1||v_2|\sin(\theta)$. 
From there, we have two cases:
\begin{itemize}[leftmargin=3.5mm]
\item If $|v_1| = 0$ or $|v_2| = 0$, then knowing $|v_1|$ and $|v_2|$ is enough because $ \begin{pmatrix} |v_1|\\ |v_2| \end{pmatrix}$ is the same as $\vb$ up to a global phase. Thus, there is no need to compute $\theta$.
 \item If $|v_1||v_2|>0$ then we can derive $\cos(\theta)$ and $\sin(\theta)$ from the above-defined quantities and get $\theta$. Thus we know all parameters of $\vb$.
\end{itemize}

Let us now assume that the state recovery is possible for $n_{qb}-1$ qubits, i.e. there is a function $f_{n_{qb}-1}$ such that for a vector $\wb$ with $2^{n_{qb}-1}$ elements $f_{n_{qb}-1}\left(\left|\Av_{t}(n_{qb}-1)\wb\right|^2\right)$ is equal to $\wb$ up to a global phase. Let $\vb$ be a $d=2^{n_{qb}}$ element vector (it does not have to be unit-norm). We split $\vb$ into two $2^{n_{qb}-1}$ element vectors $\wb_1$ and $\wb_2$: $\vb = \begin{bmatrix} \wb_1\\\wb_2 \end{bmatrix}$.
Let us show how $\vb$ can be recovered up to a global phase from $|\Av_{t}(n_{qb}) \vb|^2$ using the fact that $\wb_1$ and $\wb_2$ can be recovered form $|\Av_{t}(n_{qb}-1) \wb_1|^2$ and $|\Av_{t}(n_{qb}-1) \wb_2|^2$ up to global phases using $f_{n_{qb}-1}$.
We start by comparing $\Av_{t}(n_{qb}-1)$ to $\Av_{t}(n_{qb})$:

$\Av_{t}(n_{qb}-1) =  \begin{bmatrix}\Ev_{s_1}^*\\ \vdots \\ \Ev_{s_{2n_{qb}-1}}^*\end{bmatrix}$%P_{s_2}^T, P_{s_3}^T, ..., P_{s_{2n_{qb}-2}}^T

\noindent with (\ref{eqn:Atall}) giving the values of the strings $s_1, ..., s_{2n_{qb}-1}$. We can also notice that:

\begin{equation}
\label{eqn:i1}
\Av_{t}(n_{qb}) =  \begin{bmatrix}\Ev_{Zs_1}^*\\ \vdots \\ \Ev_{Zs_{2n_{qb}-1}}^* \\ \Ev_{X...X}^* \\ \Ev_{YX...X}^*\end{bmatrix}
\end{equation}

\noindent where $Zs_k$ is the string made up of $Z$ followed by $s_1$.

\noindent Using the definition of $\Ev$ in Section \ref{section:Pauli}, we have:

\begin{equation}
\label{eqn:i2}
\Ev_{Zs_k}^* = \Ev_Z^* \otimes \Ev_{s_k}^* = \begin{bmatrix}  \Ev_{s_k}^* & \sbv{0} \\ \sbv{0} & \Ev_{s_k}^* \end{bmatrix} \forall k .
\end{equation}

\noindent Let $k$ be an integer ranging from $1$ to $2n_{qb}-1$, from (\ref{eqn:i1}) and (\ref{eqn:i2}), we have:
\begin{equation}
\label{eqn:receq1}
|\Av_{t}(n_{qb}) \vb|^2_{i_k} = \bigg|\begin{bmatrix}  \Ev_{s_k}^* & \sbv{0} \\ \sbv{0} & \Ev_{s_k}^* \end{bmatrix} \begin{bmatrix} \wb_1\\\wb_2 \end{bmatrix} \bigg| ^2 =  \begin{bmatrix} |\Ev_{s_k}^* \wb_1|^2 \\ |\Ev_{s_k}^* \wb_2|^2 \end{bmatrix}
\end{equation}
\noindent where $|\Av_{t}(n_{qb}) \vb|^2_{i_k}$ is the vector that contains the elements of $|\Av_{t}(n_{qb}) \vb|^2$ indexed between $(k-1)d+1$ and $kd$. And using the same notation for $|\Av_{t}(n_{qb}-1) \wb_l|^2$ with $l$ being either $1$ or $2$, we have
\begin{equation}
\label{eqn:receq2}
|\Av_{t}(n_{qb}-1) \wb_l|^2_{i_k} = |\Ev_{s_k}^* \wb_l|^2 .
\end{equation}

From (\ref{eqn:receq2}) and (\ref{eqn:receq1}), we see that all the elements of $|\Av_{t}(n_{qb}-1) \wb_l|^2_{i_k}$ are in $|\Av_{t}(n_{qb}) \vb|^2_{i_k} \  \forall k  \in \{1, ..., 2n_{qb}-1\}$. Since $|\Av_{t}(n_{qb}-1) \wb_l|^2_{i_k} \ \forall k  \in \{1, ..., 2n_{qb}-1\}$ spans all the vector $|\Av_{t}(n_{qb}-1) \wb_l|^2$ we have shown that $|\Av_{t}(n_{qb}-1) \wb_l|^2$ is known from part of the measurements ($|\Av_{t}(n_{qb}) \vb|^2$) for $l=1$ and $l=2$.

Using the induction hypothesis we can apply $f_{n_{qb}-1}$ to the known quantities $|\Av_{t}(n_{qb}-1) \wb_1|^2$ and $|\Av_{t}(n_{qb}-1) \wb_2|^2$ in order to get $\wb_1$ and $\wb_2$ up to global phases. Let us call our estimates $\widehat{\wb_1}$ and $\widehat{\wb_2}$, $\wb_1= e^{i\theta_1}\widehat{\wb_1}$ and $\wb_2= e^{i\theta_2}\widehat{\wb_2} $. We now only need to know $\theta_2-\theta_1$ in order to know $\vb$ up to a global phase. Let us get $\theta_2-\theta_1$ from the last $2d$ elements of $|\Av_{t}(n_{qb}) \vb|^2$. We define $\sbm{L_m}$ as the column vector containing those last $2d$ elements 

\noindent \scalebox{0.98}{$\sbm{L_m}=\bigg| \begin{bmatrix} \Ev_{XX...X}^* \\ \Ev_{YX...X}^* \end{bmatrix} \begin{bmatrix} \wb_1 \\ \wb_2 \end{bmatrix} \bigg| ^2= \bigg| \begin{bmatrix} \Ev_X^* \otimes \Ev_{X...X}^*  \\ \Ev_Y^* \otimes  \Ev_{X...X}^* \end{bmatrix} \begin{bmatrix} \wb_1 \\ \wb_2 \end{bmatrix} \bigg| ^2$} %\bigotimes^{n_{qb}-1}_{m=1}\Ev_X

\noindent where on the left-hand side the strings $XX...X$, $YX...X$ have $n_{qb}$ characters and on the right-hand side $X...X$ have $n_{qb}-1$ characters. By replacing $\Ev_X$ and $\Ev_Y$ by their values of Section \ref{section:Pauli} and calculating the tensor products, we get 

$ \begin{matrix} \sbm{L_m} &= &\Bigg| \frac{1}{\sqrt{2}} \begin{bmatrix} \Ev_{X...X}^*\wb_1+\Ev_{X...X}^*\wb_2 \\ \Ev_{X...X}^*\wb_1-\Ev_{X...X}^*\wb_2\\ \Ev_{X...X}^*\wb_1-i \Ev_{X...X}^*\wb_2\\ \Ev_{X...X}^*\wb_1+i \Ev_{X...X}^*\wb_2 \end{bmatrix}  \Bigg| ^2 \\
&= &\frac{1}{2} \Bigg| \begin{bmatrix} \Ev_{X...X}^*\widehat{\wb_1}e^{i\theta_1}+\Ev_{X...X}^*\widehat{\wb_2}e^{i\theta_2}  \\ \Ev_{X...X}^*\widehat{\wb_1}e^{i\theta_1}-\Ev_{X...X}^*\widehat{\wb_2}e^{i\theta_2} \\ \Ev_{X...X}^*\widehat{\wb_1}e^{i\theta_1}-i \Ev_{X...X}^*\widehat{\wb_2}e^{i\theta_2} \\ \Ev_{X...X}^*\widehat{\wb_1}e^{i\theta_1}+i \Ev_{X...X}^*\widehat{\wb_2}e^{i\theta_2}  \end{bmatrix}  \Bigg| ^2. \end{matrix} $

\noindent Let us introduce the following notations % \scalebox{0.95}{}
\begin{equation}
\label{eqn:defd}
\begin{split} &\mb = \frac{1}{2}|\Ev_{X...X}^*\widehat{\wb_1}|^2+\frac{1}{2}|\Ev_{X...X}^*\widehat{\wb_2}|^2 \\
&{\sbm{d_c}} = \overline{\Ev_{X...X}^*\widehat{\wb_1}} * \Ev_{X...X}^*\widehat{\wb_2}\\
&\db(\theta) = \cos(\theta) Re(\sbm{d_c})-\sin(\theta) Im(\sbm{d_c})\end{split} \end{equation}
\noindent where $*$ again represents the element-wise product between two vectors and $\bar{.}$ is the conjugate. 
$\widehat{\wb_1}$ and $\widehat{\wb_2}$ are known quantities (from $|\Av_{t}(n_{qb}) \vb|^2$) so $\mb$ and $\sbm{d_c}$ are known and $\db(\theta)$ can be computed for any $\theta \in [0, 2\pi]$.
Let us rewrite $\sbm{L_m}$ as a function of $(\theta_2-\theta_1)$ using those quantities 

\begin{equation}
\label{eqn:L_m2}
\sbm{L_m}(\theta_2-\theta_1) = \begin{bmatrix} \mb+\db(\theta_2-\theta_1) \\ \mb-\db(\theta_2-\theta_1) \\ \mb+\db(\theta_2-\theta_1-\pi/2) \\ \mb-\db(\theta_2-\theta_1-\pi/2) \end{bmatrix} .
\end{equation}

We aim at deriving $\theta_2-\theta_1$ from $\sbm{L_m}$ (which is known from the measurements). We first notice from the definition of $\db(\theta)$ in (\ref{eqn:defd}) that if $\sbm{d_c}$ is $0$ on every component then $\db(\theta_2-\theta_1)$ is also $0$ on every component (which means it does not depend on $\theta_2-\theta_1$) and $\sbm{L_m}$ is simply $\mv$ repeated 4 times (see (\ref{eqn:L_m2})). Therefore recovering $\theta_2-\theta_1$ (and $\vb$) from $\sbm{L_m}$ is impossible. However, we hereafter show that this is the only case when $\theta_2-\theta_1$ cannot be recovered from $\sbm{L_m}$. And the ensemble of $\vb$ which make this occur has zero measure.

Let us assume that at least a single element of $\sbm{d_c}$ is not zero, let us call $k$ its index, $d_k$ the corresponding non-zero element (we take the element which has the highest modulus), and call $d_k(\theta_2-\theta_1)$ and $m_k$ the $k$-th elements of $\db(\theta_2-\theta_1)$ and $\mb$ respectively. Then all we need is the $k$-th and $(k+d)$-th elements of $\sbm{L_m}$ whose expressions are $m_k+\cos(\theta_2-\theta_1)Re(d_k)-\sin(\theta_2-\theta_1)Im(d_k)$ and $m_k+\sin(\theta_2-\theta_1)Re(d_k)+\cos(\theta_2-\theta_1)Im(d_k)$. Those known elements can be put in a column vector and re-written as: 

\begin{equation}
\label{eqn:m2}\begin{pmatrix}Re(d_k) & -Im(d_k)\\
Im(d_k) & Re(d_k)\end{pmatrix} \begin{pmatrix} \cos(\theta_2-\theta_1)\\ \sin(\theta_2-\theta_1)\end{pmatrix}.
\end{equation}

The $2\times 2$ matrix on the left-hand side is known (since $d_k$ is known) and invertible (since its determinant is $|d_k|^2>0$). Therefore $\theta_2-\theta_1$ can be recovered (because we have its sine and cosine) from 2 elements of $\sbm{L_m}$ (so two probabilities).

We could stop there and get an estimate $\theta_d$ of $\theta_2-\theta_1$ that is computed using two elements of $\sbm{L_m}$. But, in practice the sample probabilities give an imperfect estimate of $\sbm{L_m}$ which we call $\widehat{\sbm{L_m}}$. In order to be robust to the errors, we aim to find the angle $\widehat{\theta_2-\theta_1}$ that minimizes $||\sbm{L_m}(\theta_2-\theta_1)-\widehat{\sbm{L_m}}||$, this way we use all sample probabilities and not just two. We use a quasi-Newton BFGS algorithm \cite{BFGS} (implemented with fminunc in the Matlab numerical software) initialized at $\theta_d$, the optimization stops when the step is smaller than $10^{-30}$. Technically with this optimization, the algorithm is no longer closed-form but, since it involves a single parameter, it is really fast, and improves the performances quite significantly so we choose to perform it anyway. If the readers want a real closed-form algorithm, they can use $\theta_d$ instead of computing $\widehat{\theta_2-\theta_1}$, or use a closed-form optimization algorithm with a fixed number of steps to compute $\widehat{\theta_2-\theta_1}$.

Let us now take a step back and summarize what we have proved in this section:
\begin{itemize}[leftmargin=3.5mm]
\item Recovering the state (up to a global phase) from the measurements is possible for $n_{qb}=1$.
\item Assuming it is possible for $n_{qb}-1$ we showed it is also possible for $n_{qb}$ unless the state is in an ensemble of zero measure. 
\end{itemize}
Using those previous two results, we can construct a recursive algorithm that recovers $\vb$ from the measurements. It will work except on the union of a finite number of failure sets of zero measure which would also be of zero measure. The estimate given by this recursive algorithm will be called $\widehat{\vb}_{rec}$.

\subsection{Discussion about the number of probabilities used}
\label{section:n_mes}
The recursive algorithm of the previous section calls itself twice for each reduction of the number of qubits by $1$. This means that for $n_{qb}$, it is called once with $n_{qb}$ qubits, twice with $n_{qb}-1$ qubits, ..., $2^{n_{qb}-1}$ times with $1$ qubit. 

For $1$ qubit, the state is recovered using (\ref{eqn:A1}) which involve 6 probabilities, among which only 4 are required (we could obtain the same result without using the fourth and sixth elements of $|\sbv{A_t}(1)\vb|^2$).

For $q>1$ qubit before calling the recursive function with one fewer qubit, we compute $\theta_2-\theta_1$ using (\ref{eqn:L_m2}). This involves $2\times 2^q$  probabilities among which only $2$ are strictly required for the first estimate $\theta_d$.

The minimum number of needed probabilities is $4 \times 2^{n_{qb}-1} + 2\sum_{q=2}^{n_{qb}}2^{n_{qb}-q}=2d+2(2^{n_{qb}-1}-1)=3d-2$. Furthermore, if we take into account the fact that $\vb$ has unit norm, then one of the probabilities along the $Z$ axis (which are all used) becomes redundant, and this number becomes $3(d-1)$.

In practice all probabilities are used in order to minimize the impact of the statistical error on the probabilities. But if we wanted to remove rows from ${\Av_{t}}$ in (\ref{eqn:Atall}) and only keep $3(d-1)$ of them, we could still achieve QST. However, this is a bad idea because we would no longer have a concatenation of $d$-outcome parallel measurements. And in practice the final estimate of the state would be less robust to the errors on the sample probabilities and the quantum setup would not be any easier to put in place, as the estimation of the  $3(d-1)$ probabilities to be kept requires all $2n_{qb}+1$ measurements to be performed anyway.

\subsection{Comparison with the literature}
\label{section:compLit2}

Let us sum up the main features of our second QST algorithm:
\begin{itemize}
\item It uses $(2n_{qb}+1)d$ probabilities that can be obtained by averaging the results of $2n_{qb}+1$ parallel unentangled measurements.
\item The measurements are injective outside a known failure set with zero measure.
\item The algorithm that reconstructs the state is explicit.
\end{itemize}
Those features are very similar to those of Goyeneche et al. \cite{Goyeneche}. The advantage of our method it that the measurements it uses are unentangled. Its drawback is that it requires $2 n_{qb}+1$ measurements which is more than $4$ (except for the trivial case $n_{qb} = 1$). That is the price to pay for using only unentangled measurements, we could not find a simple closed-form algorithm that works with fewer types of unentangled measurements. 
The more general compressed sensing approach of \cite{CS3} requires $O(r d \log(d)^2)$ probabilities to estimate the state where $r$, the rank of the density matrix, is $1$ in the case of a pure state. Those probabilities could be  obtained by averaging the results of $O(\log(d)^2)$ different unentangled measurements. We do better here since we only use $2n_{qb}+1=O(\log(d))$ measurements. We also have the advantage of providing a closed-form algorithm contrary to the method of \cite{CS3} which is very general (works for mixed states and any kind of measurement), but uses an optimization algorithm and provides no proof of injectivity.

\section{Likelihood Maximization}
\label{section:ML}
\subsection{Main idea}
\label{section:intro_ml}
Sections \ref{section:petitA} and \ref{section:grandA} give us estimates of the state $\vb$, denoted as $\widehat{\vb}_{pc}$ and $\widehat{\vb}_{rec}$ respectively. $\widehat{\vb}_{pc}$ is the solution of the QST problem with one constraint ($rank(\Uv)=1$) relaxed, so it can be inaccurate even in the absence of errors in the sample probabilities. The algorithm of Section \ref{section:pinj} that computes  $\widehat{\vb}_{rec}$ is also imperfect. It relies heavily on the measurements along $Z...Z$, $Z..ZX$ and $Z..ZY$ (used $2^{n_{qb}-1}$ times for one qubit at the end of the recursive tree to compute all the moduli and half the phases differences) and it almost does not use the measurements along $X...X$ and $YX...X$ (used only once to compute one phase difference ($\theta_2-\theta_1$) with (\ref{eqn:L_m2})). Each of those last two measurements contains as much information on $\vb$ as the measurements along $Z...Z$, but the former are barely used.

Therefore the estimation methods of Section \ref{section:petitA} and \ref{section:grandA} are hereafter supplemented by a final tuning to make them more precise. To this end, we take a maximum likelihood (ML) approach:
\begin{equation}
\label{eqn:MV}
(\widehat{\xb}, \widehat{\yb}) = \underset{\xb, \yb \text{  s.t. } ||\xb||_2+||\yb||_2<1}{\arg\min} \mathscr{L}_{(\xb, \yb)}(\widehat{\pb})
\end{equation}

\noindent where $\widehat{\pb}$ is the vector that contains sample probabilities and $\mathscr{L}_{(\xb, \yb)}(\widehat{\pb})$ is to be understood as the negative log-likelihood of measuring the sample probabilities $\widehat{\pb}$ if the true state is $\vb(\xb, \yb)$, with $\xb$ and $\yb$ defined hereafter. In the whole paper, whenever we write ``negative log-likelihood"  (or $\mathscr{L}$) we mean ``opposite of the log-likelihood up to additive and positive multiplicative constants". These constants will not matter as the negative log-likelihood will be minimized. The vector $\vb(\xb, \yb)$ with respect to which $\mathscr{L}$ will be minimized is defined as:

\noindent \scalebox{0.9}[0.9]{$
\vb(\xb, \yb)= [\sqrt{1-||\xb||_2^2-||\yb||_2^2}, x_1+iy_1, ...,  x_{d-1}+iy_{d-1}]^T.
$}

$\xb$ and $\yb$ are $d-1$ element vectors representing the real and imaginary parts of the last elements of $\vb$. 
The constraint in (\ref{eqn:MV}) is $r^2<1$ (with $r=\sqrt{||\xb||_2^2+||\yb||_2^2}$) and not $r^2\leq 1$ because optimization is easier on an open set). We mitigate the effect of this imperfect constraint by permuting the first component of $\vb$ and the component of $\vb$ with the highest modulus at the initial point of the optimization. Thus, we ensure that $r^2$ is not going to be close to $1$ unless the initial point was way off. The sample probabilities and the columns of $\Av$ are permuted in the same way. Those change are limited to the optimization algorithm. 

Since the optimization set is open we can change the variables in order to remove the constraint altogether:

 $\xb'=\frac{tan(\frac{\pi}{2} r)}{r}\xb$ and $\xb=\frac{\frac{2}{\pi}atan( r')}{r'}\xb'$

 $\yb'=\frac{tan(\frac{\pi}{2} r)}{r}\yb$ and $\yb=\frac{\frac{2}{\pi}atan( r')}{r'}\yb'$

\noindent (with $r'=\sqrt{||\xb'||_2^2+||\yb'||_2^2}$). The new optimization problem on $\xb'$and $\yb'$ does not have any constraint, as when $r'$ spans the whole space $r$ remains strictly smaller than $1$. Eq. (\ref{eqn:MV}) is therefore replaced by

\begin{equation}
\label{eqn:MV2}
(\widehat{\xb'}, \widehat{\yb'}) = \underset{\xb', \yb'}{\arg\min} \mathscr{L}_{(\xb', \yb')}(\widehat{\pb}).
\end{equation}

In order to solve (\ref{eqn:MV2}) we again use the BFGS algorithm where the analytical expressions of the gradients are provided. The algorithm stops when the norm of the optimization step is smaller than $10^{-30}$. Like in most non-convex optimization methods, we need a good initialization point, we use either $\widehat{\vb}_{pc}$ or $\widehat{\vb}_{rec}$. The most likely $\vb$ is $\widehat{\vb}_{ml}= \vb(\widehat{\xb'}, \widehat{\yb'})$, with $\widehat{\xb'}, \widehat{\yb'}$ defined in (\ref{eqn:MV2}). %From now on we omit all the $'$, $x'$ will be written as $x$ and $y'$ as $y$.

All that remains now is to define the expression of the negative log-likelihood $\mathscr{L}$ with respect to $\vb$. In the following 2 subsections we will give 2 expressions for the normalized log-likelihood: $\mathscr{L}_{(\xb', \yb')}^{exact}(\widehat{\pb})$ and $\mathscr{L}_{(\xb', \yb')}^{gauss}(\widehat{\pb})$.
\subsection{Exact likelihood}
\label{section:bin_ml}

In \cite{ML} the formula for the likelihood of a multi-output quantum measurement is given (albeit for a mixed state represented by $\sbv{\rho}$ which we would have to replace by $\vb\vb^*$). It boils down to:
\begin{equation}
\label{eqn:V1}
\mathscr{L}_{(\xb', \yb')}^{exact}(\widehat{\pb}) = -\sum_{k=1}^{n_{prob}} n_k log\big((|\Av \vb(\xb', \yb')|^2)_k\big) .
\end{equation}
\noindent $\left(|\Av \vb(\xb', \yb')|^2\right)_k$ is the $k$-th element of $|\Av \vb(\xb', \yb')|^2$, $\Av$ is the measurement matrix, either $\Av_{s}$ or $\Av_{t}$, $n_k$ is the number of times the $k$-th outcome occurred i.e. the $k$-th element of $\widehat{\pb}$ (either $\widehat{\sbm{p_s}}$ or $\widehat{\sbm{p_t}}$) multiplied by the number of times the measurement is repeated, and $n_{prob}$ is the number of rows of $\Av$. %where $\Av$ is the measurement matrix, so either $\Av_{s}$ or $\Av_{t}$, 

In order to get to this result we must consider the measurement counts as the realizations of a multinomial random variable. This is not an approximation, this is why we call this likelihood ``exact".
\subsection{Gaussian approximation}
\label{section:gauss_ml}
In this subsection, we use the central limit theorem to approximate the scaled sample probabilities as the realization of a multivariate normal distribution. It is appropriate as the vector $\widehat{\pb}$ whose likelihood we want to compute is the average of independent realizations of the same random variable. Its expected value is the vector of theoretical probabilities $\pb(\xb', \yb')$ that depends on the state. Let us define $\sbm{\sbm{\varepsilon}} (\widehat{\pb}, \xb', \yb') =  \widehat{\pb}-\pb(\xb', \yb')$ and $\barbelow{\sbm{\varepsilon}} (\widehat{\pb}, \xb', \yb')$ is $\sbm{\varepsilon} (\widehat{\pb}, \xb', \yb') $ with the last element removed (no information is lost as the sum of the elements of $\sbm{\varepsilon} (\widehat{\pb}, \xb', \yb')$ is 0). In Appendix \ref{section:cov}, we show that if $N$ is the number of times the measurements have been averaged, then $\sqrt{N} \barbelow{\sbm{\varepsilon}} (\widehat{\pb}, \xb', \yb')$ asymptotically ($N\to +\infty$) follows a zero-mean multivariate normal distribution. Its covariance matrix $\sbv{\Sigma}$ is computed in Appendix \ref{section:cov}. $\sbv{\Sigma}$ depends on the theoretical probabilities, we need to remove this dependency. With that in mind, we get to the following approximation for the negative log-likelihood:
\begin{equation}
\mathscr{L}_{(\xb', \yb')}^{gauss}(\widehat{\pb})= N \barbelow{\sbm{\varepsilon}} (\widehat{\pb}, \xb', \yb')^T \widetilde{\sbv{\Sigma}}^{-1} \barbelow{\sbm{\varepsilon}} (\widehat{\pb}, \xb', \yb')
\end{equation}
\noindent where $\widetilde{\sbv{\Sigma}}^{-1}$ is an approximation of the covariance matrix that uses $\widetilde{\pb} = \frac{\widehat{\pb}+\frac{5}{N}}{1+\frac{5d}{N}}$ as a regularized approximation of $\pb$, this is justified in Appendix \ref{section:cov}.
Appendix \ref{section:cov} also shows that this equation boils down to 
\begin{equation}
\mathscr{L}_{(\xb', \yb')}^{gauss}(\widehat{\pb})= N \sum_{k=1}^{d}{\frac{\varepsilon_k (\widehat{\pb}, \xb', \yb')^2}{\widetilde{p_k}}}.
\end{equation}

This log-likelihood is the result of two approximations that are true only when $N\to +\infty$: we approximated $\barbelow{\sbm{\varepsilon}} (\widehat{\pb}, \xb', \yb')$ as the realization of a Gaussian random vector and we used an approximation for $\sbv{\Sigma}$. In practice, the resulting approximation is smoother and easier to minimize than $\mathscr{L}_{(\xb', \yb')}^{exact}(\widehat{\pb})$ if the initialization point is not good enough (as will be shown in Section \ref{section:err_init}). However, with a good initialization, the state that minimizes $\mathscr{L}_{(\xb', \yb')}^{exact}(\widehat{\pb})$ should be closer to the true state than the one that minimizes $\mathscr{L}_{(\xb', \yb')}^{gauss}(\widehat{\pb})$. The smaller $N$, the starker the difference. This will be shown in Section \ref{section:Lcomp}.
\subsection{Mixed minimization}
\label{section:mixed}
As stated above $\mathscr{L}_{(\xb', \yb')}^{gauss}$ is supposed to be easier to minimize but the minimum of $\mathscr{L}_{(\xb', \yb')}^{exact}$ is supposed to be a better estimate. A good way to combine the two advantages is to start the optimization process by minimizing $\mathscr{L}_{(\xb', \yb')}^{gauss}$  and finish it by minimizing $\mathscr{L}_{(\xb', \yb')}^{exact}$. In practice, we here again run the BFGS algorithm on $\mathscr{L}_{(\xb', \yb')}^{gauss}$ for 100 iterations starting from the initialization point of Sections \ref{section:petitA} or \ref{section:grandA}, this yields $\widehat{\vb}_{inter}$. And then we run the BFGS algorithm on $\mathscr{L}_{(\xb', \yb')}^{exact}$ starting from $\widehat{\vb}_{inter}$ and stopping only once a local (hopefully global) minimum has been found.
\section{Numerical Results}
\label{section:plot}
\subsection{Performances of the two initialization algorithms}
\label{section:intPerf}
Sections \ref{section:petitA} and \ref{section:grandA} detail two methods to perform QST which are used for initialization of ML algorithms. The current section aims at estimating the precision of those methods and comparing them whenever possible. The recursive algorithm of Section \ref{section:grandA} only works for a specific set of measurement types but is explicit and does not require an undefined number of iterations to converge contrary to PhaseCut defined in Section \ref{section:petitA}. We only explained PhaseCut for the setup with 4 different measurement types described in Section \ref{section:mess}, but it can be applied to any types of measurements. In particular we could apply it to the setup with $2n_{qb}+1$ measurement types of Section  \ref{section:mess2}. In the current section, we test both PhaseCut and the recursive algorithm on 50 randomly generated 7-qubit pure states. The two sets of measurement types of Sections \ref{section:mess} and \ref{section:mess2} are considered. They contain respectively $4$ and $2\times7+1=15$ measurement types. We test those algorithms with 2 different fixed numbers of total measurements $N_C$: 5 000 and 500 000. Thus each one of the 4 measurement types of the setup of Section \ref{section:mess} is performed either $N_C = 1 250$ or $N_C=125 000$ times and each one of the 15 measurement types of the setup of Section \ref{section:mess} is performed either $N_C = 333$ or $N_C = 33 333$ times. 

The metric used in order to quantify the proximity of $\widehat{\vb}$ to the actual vector $\vb$ up to a phase factor is $\mu = ||\vb-\widehat{\vb}.e^{-i\xi}||_2$ with $\xi$ the angle that minimizes our metric: $e^{i\xi} = \frac{\vb^*\widehat{\vb}}{|\vb^*\widehat{\vb}|}$. We call $\mu$ this error in the rest of the paper. $\mu$ is maximal for orthogonal states (it is then $\sqrt{2}$), and minimal for states that differ by a global phase (is is then $0$). A more widely used metric in the literature is the fidelity (see Section 9.2.2 in \cite{bookNC}) $f = |\vb^*\widehat{\vb}|$. It can be shown that $f = (1-\frac{\mu^2}{2})$. We do not use the fidelity because it can push some interesting values too close to 1. 

Fig. 1 shows the error of $\widehat{\vb}_{pc}$ obtained by using PhaseCut with 100 to 100 000 iterations for the two setups (4 and 15 measurement types). With 15 (and not with 4) measurement types, the recursive algorithm can be implemented. We display biggest and smallest errors of $\widehat{\vb}_{rec}$ obtained with the recursive algorithm with horizontal bold green and red lines respectively. The recursive algorithm is performed in a fixed number of steps, this is why we plot the error on horizontal lines and not on a curve with respect to a number of iterations.

The aim of this simulation is to see how many iterations of PhaseCut are required to get a good estimate of the state and to compare the performances of the recursive algorithm with those of the more versatile PhaseCut. 
\begin{figure}[h]
\hspace{-0mm}
    \includegraphics[width=8cm]{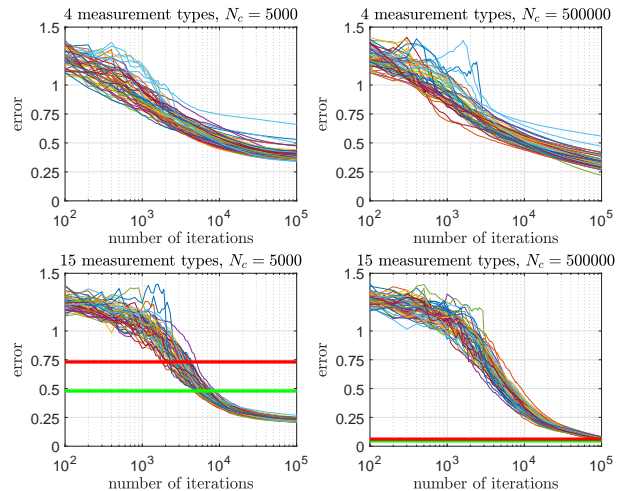}
    \caption{Initialization algorithms. The bold red and green horizontal lines are the worst and best errors for the recursive algorithm on the 50 random states (only available with 15 measurement types). The other curves represent the evolution of the error on the PhaseCut estimates with the 50 states.}
	\label{fig:eps}
\vspace{-1mm}
\end{figure}

With enough iterations ($\sim 10^4$ for $N_c = 5000$ and $\sim 10^5$ for $N_c = 500000$) PhaseCut is more precise than the recursive algorithm on the setups on which they can both be implemented, but it takes way more time. Each iteration of PhaseCut is costly, because we are working on an $n_{prob}\times n_{prob}$ matrix. With Matlab, on a 2.11 GHz 4-core processor with 32 Go RAM, each iteration of PhaseCut takes around  $4$ ms for the setup with 4 types of measurements and around $45$ ms for the setup with 15 types of measurements. In that same 15 measurement type setup, the recursive algorithm takes $200$ ms. This is way faster than PhaseCut which runs in minutes, as it requires thousands of iterations.
%In Section \ref{section:Lconv} we will see what precision on the initialization is required. From there we can choose the number of iterations of PhaseCut and see whether the recursive algorithm's performances are good enough.
\subsection{Likelihood estimator comparison}
\label{section:Lcomp}
In Section \ref{section:ML} we defined two likelihood estimators, based on the likelihood maximization. The first one minimizes the true negative log-likelihood $\mathscr{L}_{(\xb', \yb')}^{exact}$ and the other minimizes a version of the negative log-likelihood that is supposed to be smoother, namely $\mathscr{L}_{(\xb', \yb')}^{gauss}$. We know that $\mathscr{L}_{(\xb', \yb')}^{gauss}$ is an approximation of the likelihood that is accurate only if the number of measurements per measurement type is high enough. Therefore we expect the global minimum of $\mathscr{L}_{(\xb', \yb')}^{gauss}$ to be a worse estimator than the global minimum of $\mathscr{L}_{(\xb', \yb')}^{exact}$ for a limited number of measurements. In order to check whether this is true and quantify the difference, we compute the errors on both estimators when they are initialized at the true state $\vb$. Doing this ignores the error on the initialization point (to which the regularized Gaussian estimate is supposed to be robust). We also compute the error for the mixed algorithm which starts by minimizing $\mathscr{L}_{(\xb', \yb')}^{gauss}$ and then minimizes $\mathscr{L}_{(\xb', \yb')}^{exact}$.  These 3 types of errors are computed with 1000 random initial states on the four setups described in Section \ref{section:intPerf} with 4 or 15 measurement types and 5000 or 500 000 total measurements. For each of the four setups, the empirical cumulative density function (empirical cdf) is computed on the 1000 errors associated with the initial states, those cdf are shown in Fig. 2. 

\begin{figure}[h]
    \centering
    \includegraphics[width=8cm]{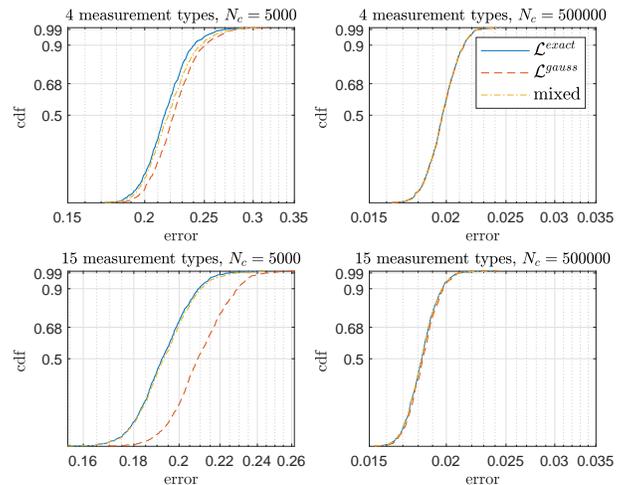}
    \caption{empirical cdf of the errors of the 3 maximum likelihood estimators}
	\label{fig:eps}
\vspace{-1mm}
\end{figure}
As predicted the error is larger with the Gaussian estimate of the likelihood, and the difference decreases when the number of measurements per measurement type increases. 

The performance of the mixed minimization algorithm is very close to that of the estimator that minimizes $\mathscr{L}_{(\xb', \yb')}^{exact}$. There can be small differences however. Its turns out that they sometimes converge toward close but different minima. This is due to the fact that the small error made by the first 100 iterations  of the mixed algorithm (during which $\mathscr{L}_{(\xb', \yb')}^{gauss}$ is minimized) can be enough to affect the final result. 

The differences between the 3 estimators are only noticeable for $N_c = 5000$ with 15 and 4 different measurements (so 333 or 1250 measurements per measurement type). 
\subsection{Convergence of the likelihood estimators}
\label{section:err_init}
In the current section, we intend to see what precision on the initial state is required to make sure that the likelihood optimization algorithm converges towards a reasonable solution, and compare the robustness of the three ML estimates. We compare the rates of divergence (denoted as $\delta$ and defined below) of the algorithms that minimize $\mathscr{L}_{(\xb', \yb')}^{exact}$ and $\mathscr{L}_{(\xb', \yb')}^{gauss}$ as well as the mixed algorithm. 1000 random states $\vb$ to be estimated are considered with 1000 associated initial states of ML algorithms that have an initialization error $\mu$ linearly varied from $0$ to $\sqrt{2}$ (as stated above $\sqrt{2}$ is the highest possible value for $\mu$, it is reached if the two states are orthogonal). Let us denote as $\{ \mu_i \}_{i\in\{1, ..., 1000\}}$ the 1000 values of this initial error on states $\vb$ and define $\left\{b_i^{algo}, i\leq1000, algo \in \{exact, Gauss, mixed\}\right\}$ where $b_i^{algo}$ is $-1$ if the $algo$ algorithm converges towards the same minimum with the $\mu_i$ initialization error and with no error and $+1$ if it converges toward a different minimum. We say that those two minima are the same if the error $\mu$ between the two is smaller than one percent of the error between the first one (initialized without error) and the true state vector.%$\{b_i^{exact}, b_i^{gauss}, b_i^{mixed}\}_i$

For each of the 3 algorithms, we then define the rate of divergence $\delta_{algo}(\mu)$ associated with a given error $\mu$. It takes all the $b_i^{algo}$ into account but gives more weight to those for which the associated $\mu_i$ is close to $\mu$:

$\delta_{algo}(\mu) = \frac{1}{2}\Big(1+\frac{\sum_{i=1}^{1000} b_i^{algo} e^{-\left(\frac{\mu-\mu_i}{\alpha}\right)^2}}{\sum_{i=1}^{1000} e^{-\left(\frac{\mu-\mu_i}{\alpha}\right)^2}} \Big)$.

Simply put, if the majority of $\mu_i$ in the vicinity of $\mu$ are associated with $b_i^{algo}$ equal to $-1$ (i.e. the algorithm converges towards the proper minimum with initialization errors around $\mu$) then, $\delta_{algo}(\mu)$ will be close to $0$. If the associated $b_i^{algo}$ are $1$ (i.e. the algorithm does not converge towards the proper minimum) then, $\delta_{algo}(\mu)$ will be close to $1$. The parameter $\alpha$ quantifies how far away from $\mu$ we look for results, we picked  $\alpha =0.1$. Fig. 3 shows the rates of divergence of the 3 algorithms in the four setups described in Section \ref{section:intPerf} with 4 or 15 measurement types and 5000 or 500 000 total measurements.
\label{section:Lconv}
\begin{figure}[h]
    \centering
    \includegraphics[width=8cm]{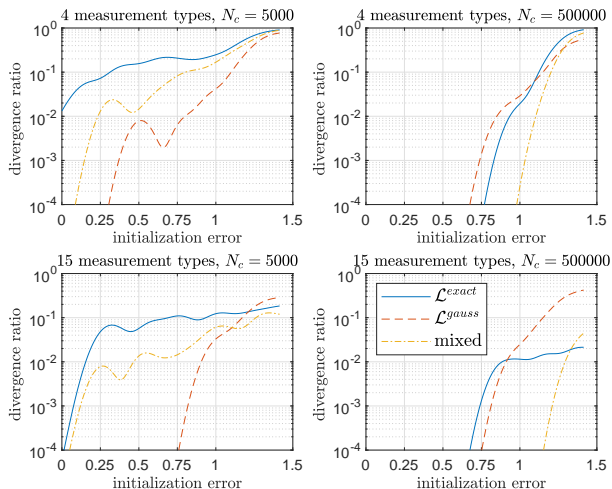}
    \caption{Convergence of the different likelihood algorithms in the presence of initialization errors.}
	\label{fig:eps}
\vspace{-1mm}
\end{figure}

The two plots on the right are of limited interest to us as the rate of divergence is always very low ($\leq 10^{-4}$) for errors lower than $0.75$. We are mostly interested in the rates of divergence for initialization errors $\mu$ smaller than $0.75$ because according to Fig. 1, the recursive algorithm always yields an estimate that corresponds to an error lower than $0.75$ and PhaseCut also does so quite quickly (for more than 5000 iterations) for every setup. 
For those errors (on the two plots on the left), the best algorithm seems to be the minimization of $\mathscr{L}_{(\xb', \yb')}^{gauss}$, indeed increased robustness to the initialization error is the whole reason why we introduced $\mathscr{L}_{(\xb', \yb')}^{gauss}$. The mixed algorithm does not quite reach the same robustness but it is certainly an improvement over the algorithm that minimizes $\mathscr{L}_{(\xb', \yb')}^{exact}$ which has the worst performances for the relevant initialization errors. We should note that the name given to $\delta$: ``rate of divergence" is a bit severe as the likelihood algorithms never diverge in practice, they simply converge toward a false local minimum that is sometimes close to the real global minimum. $\delta$ is not useless however, and Fig. 3 shows us that, generally, with either the mixed algorithm or the algorithm that minimizes $\mathscr{L}_{(\xb', \yb')}^{gauss}$, an initialization error lower $0.75$ leads to proper convergence towards the real minimum. According to Fig. 1, 5000 iterations of PhaseCut as well as the recursive algorithm generally yield an error smaller than 0.75. Therefore we choose to use the recursive algorithm when it is possible i.e. with the setup of Section \ref{section:grandA} with 15 types of measurements for 7 qubits (because it is faster than PhaseCut) and when PhaseCut has to be used (so with 4 measurement types) we only perform 5 000 iterations. We could let PhaseCut run longer but our implementation of the ML algorithm is faster.
\subsection{Global performances}
\label{section:perfs}

This section aims to test the algorithms of Sections \ref{section:petitA} and \ref{section:grandA}, fine tuned with the 3 algorithms of Section \ref{section:ML} on $n_{qb}=7$ qubits, with the four setups described in Section \ref{section:intPerf}.
For each setup, and for each version of the ML algorithm, 4 estimates of $\vb$ are computed:
\begin{itemize}[leftmargin=3.5mm]
\item The initial estimate, so $\widehat{\vb}_{pc}$ for the setup with 4 measurement types or $\widehat{\vb}_{rec}$ for the setup with 15 measurement types. It does not depend on the choice of the ML algorithm.
\item $\widehat{\vb}_{ml}$ which is the result of the likelihood optimization (minimizing either $\mathscr{L}_{(\xb', \yb')}^{exact}$ or $\mathscr{L}_{(\xb', \yb')}^{gauss}$ or both successively) initialized at the initial estimate.
\item $\widehat{\vb}_{ref}$ which is the result of the likelihood optimization initialized at the true $\vb$ (not available in practice, it should be the global maximum likelihood; if $\widehat{\vb}_{ml}=\widehat{\vb}_{ref}$ then the initial estimate was good enough). We call $\widehat{\vb}_{ref}$ the reference, it has already been defined (but not named) in Section \ref{section:Lcomp} and represented in Fig. 2.
\item And $\widehat{\vb}_{rnd}$ which is the result of the likelihood optimization initialized at a random normalized vector (if $\widehat{\vb}_{rnd}$ is not worse than $\widehat{\vb}_{ml}$, then the initial estimate was unnecessary and one can only use the maximum likelihood algorithm initialized randomly).
\end{itemize}
%Using Matlab on a laptop (2.11 GHz 4-core processor, 32 Go RAM)  for 7 qubits, the PhaseCut algorithm with 5000 iteration that we use to compute $v_{cvx}$ takes around $30s$ using a mex file, the subsequent likelihood maximization takes around 5s. With the second setup of $15$ measurements, computing $v_{end}$ with the algorithm of \ref{section:grandA} takes around $0.3s$ and the subsequent likelihood maximization takes around $20s$.

For each setup, 1000 tests are performed with 1000 randomly generated $\vb$. We compute the estimates of each $\vb$ with the different algorithms and display the empirical cumulative density function (cdf) of the errors in Fig. 4 to Fig. 6. 

The performances of the three ML algorithms are quite similar (when excluding the random initialization), but some differences can be noted:
\begin{itemize}[leftmargin=3.5mm]
\item The algorithm that minimizes $\mathscr{L}_{(\xb', \yb')}^{exact}$ is supposed to be less robust to the initialization error than the others. It is only apparent for the setup with 4 measurements and $N_c = 5000$. $\widehat{\vb}_{ml}$ is not quite as precise as $\widehat{\vb}_{ref}$.  
\item The algorithm that minimizes $\mathscr{L}_{(\xb', \yb')}^{gauss}$ does not have that problem, $\widehat{\vb}_{ml}$ and $\widehat{\vb}_{ref}$ are always indistinguishable. However the version of $\widehat{\vb}_{ref}$ computed by minimizing $\mathscr{L}_{(\xb', \yb')}^{gauss}$ is not as precise as the version that minimizes $\mathscr{L}_{(\xb', \yb')}^{exact}$. This can be seen by comparing Fig. 4 and Fig. 5 but it is more visible on Fig. 2 that represents the performances of the 3 references on a single graph.
\item The mixed algorithm seems to combine the advantages of those based on $\mathscr{L}_{(\xb', \yb')}^{gauss}$ and $\mathscr{L}_{(\xb', \yb')}^{exact}$. $\widehat{\vb}_{ml}$ is almost equal to $\widehat{\vb}_{ref}$, and $\widehat{\vb}_{ref}$ is almost as good with this mixed algorithm as with $\mathscr{L}_{(\xb', \yb')}^{exact}$ (see Fig. 2 for a clearer comparison of the two values of $\widehat{\vb}_{ref}$).
\end{itemize}
%\newpage

\begin{figure}[H]
    \centering
    \includegraphics[width=8cm]{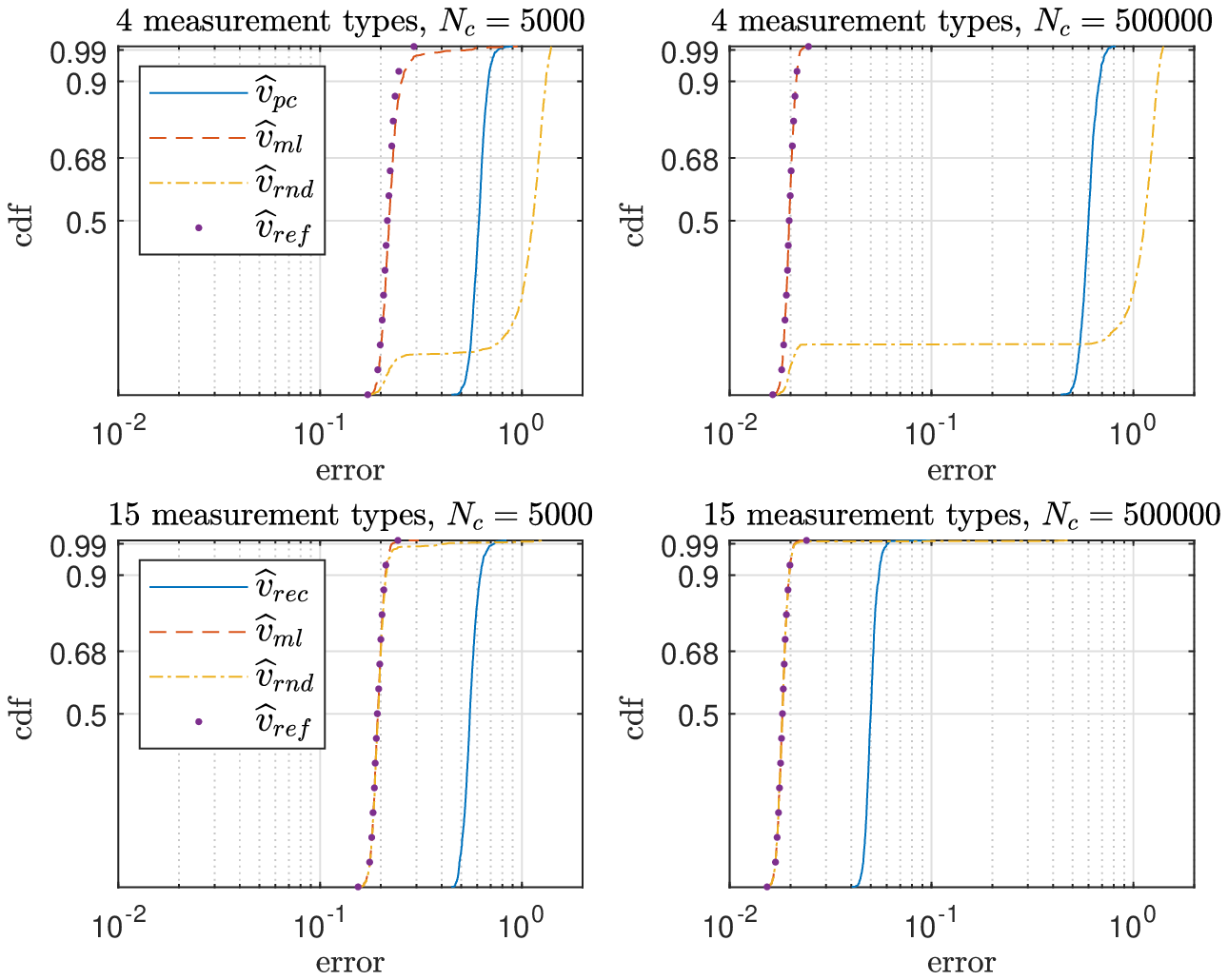}
    \vspace{-3mm}
    \caption{Empirical cdf of QST error, $\mathscr{L}_{(\xb', \yb')}^{exact}$ minimization.}
	\label{fig:eps}
\vspace{-5mm}
\end{figure}
\begin{figure}[H]
    \centering
    \includegraphics[width=8cm]{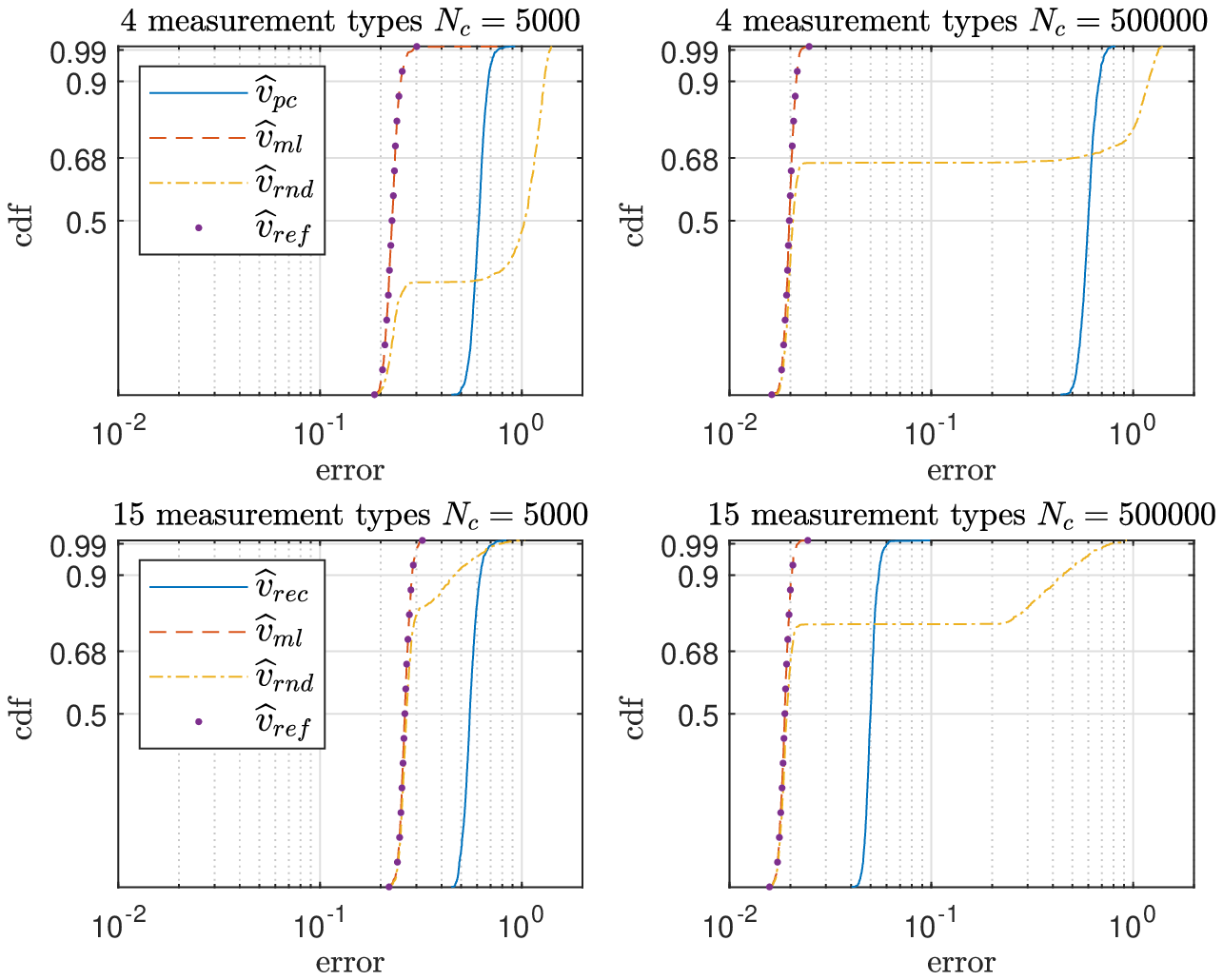}
    \vspace{-3mm}
    \caption{Empirical cdf of QST error, $\mathscr{L}_{(\xb', \yb')}^{gauss}$ minimization.}
	\label{fig:eps}
\vspace{-5mm}
\end{figure}
\begin{figure}[H]
    \centering
    \includegraphics[width=8cm]{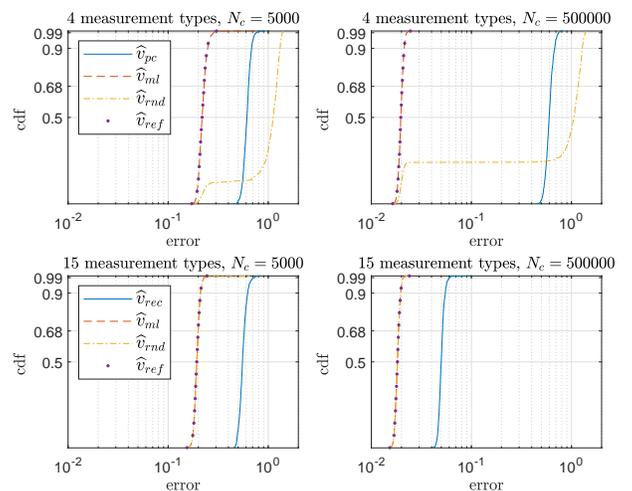}
    \vspace{-3mm}
    \caption{Empirical cdf of QST error, mixed algorithm.}
	\label{fig:eps}
\vspace{-1mm}
\end{figure}

The performances of $\widehat{\vb}_{rnd}$, the maximum likelihood estimators initialized at a random point, are interesting. With the 4 measurement type setup, it is always a much worse estimate than $\widehat{\vb}_{ml}$. But with 15 measurement types it is (almost) as good as the maximum likelihood estimators initialized at $\widehat{\vb}_{rec}$ (unless we use the $\mathscr{L}_{(\xb', \yb')}^{gauss}$ minimization). This could make us question the relevance of the recursive algorithm defined in Section \ref{section:grandA}. It would seem that the structure of the measurement matrix $\Av_{t}$ is such that the gradient descent algorithm naturally converges towards the global minimum from any initial point. However the recursive algorithm is still useful because it is very fast and speeds up the likelihood maximization (see Table 2). 

We can also compare the performances of the two initialization algorithms $\widehat{\vb}_{pc}$ or $\widehat{\vb}_{rec}$ (blue curve) with $\widehat{\vb}_{ml}$ (dashed red curve). The error on $\widehat{\vb}_{ml}$ is at least $3$ times smaller (or way less for $\widehat{\vb}_{pc}$ and $N_c =  500000$) than that of the initialization algorithms. This shows that the fine tuning with ML is very useful to reduce the error. Comparing the precision of the initialization algorithm with $\widehat{\vb}_{rnd}$ is unwise because $\widehat{\vb}_{pc}$ and $\widehat{\vb}_{rec}$ can be improved with the ML algorithm whereas $\widehat{\vb}_{rnd}$ cannot as it is a local minimum of the likelihood. Furthermore, with $N_c=5000$, $\widehat{\vb}_{pc}$ and $\widehat{\vb}_{rec}$ have a similar accuracy (respectively on 4 and 15 measurement types). And with $N_c=500 000$, $\widehat{\vb}_{rec}$ is a way better estimate than $\widehat{\vb}_{pc}$ because the PhaseCut algorithm is limited to 5000 iterations (allowing it enough iterations to converge properly would be way slower and not as accurate as the likelihood maximization).

After likelihood optimization the performances of $\widehat{\vb}_{ml}$ with 15 and 4 measurement types are comparable (with the mixed algorithm, the 15 measurement setup is slightly better). Also the final error is roughly 10 times smaller when the number of measurements is multiplied by 100. This means that for more than 5000 measurements one can extrapolate the error (and therefore its cdf), as the error is proportional to $N_c^{-1/2}$.

The fact that the recursive algorithm used to compute $\widehat{\vb}_{rec}$ has a zero measure failure set on which phase recovery is impossible (see Section \ref{section:grandA}) turns out to be a non-issue. We could have expected to see some outliers on the error of $\widehat{\vb}_{rec}$, and the $\widehat{\vb}_{ml}$ computed from it, if the randomly generated $\vb$ was close enough to the failure set. It is not the case, each one of the 1000 initial states has been successfully recovered with a reasonable error. The same is true when using PhaseCut with the 4 measurement type setup. Even though we were not able to prove the injectivity, the QST goes well in practice and there are no outliers in the error if the proper algorithms are used.

Table 1 and Table 2 give the median execution time of all the algorithms on an Intel Xeon Gold 6226R 2.9 GHz core, all the scripts ran on 1 thread on Matlab. There are no significant differences between the 3 ML algorithms when they are not initialized at random. The random initialization is never relevant, as for the 4 measurement type setup it is relatively fast (as it spares us the initialization step with PhaseCut) but inaccurate; and for the 15 measurement type setup it is always slower (sometimes way slower) that the likelihood maximization with proper initialization.
\begin{center}
%\vspace{-3mm}
\begin{table}[H]
\begin{tabular}{c c c } 
\hline
   &$N_c=5000$ &$N_c=500000$ \\ [0.25ex] 
 \hline
 PhaseCut & $16.9$ s & $17.4$ s  \\ 
 \hline
 $\mathscr{L}_{(\xb', \yb')}^{exact}$ min. from $\widehat{\vb}_{pc}$ & $11.4$ s & $8.3$ s \\ 
 \hline
 $\mathscr{L}_{(\xb', \yb')}^{exact}$ min., random init. & $22$ s & $24.7$ s \\ 
 \hline
 $\mathscr{L}_{(\xb', \yb')}^{gauss}$ min. from $\widehat{\vb}_{pc}$ & $8.1$ s & $5.2$ s \\ 
 \hline
 $\mathscr{L}_{(\xb', \yb')}^{gauss}$ min., random init. & $16.6$ s & $21.3$ s \\ 
 \hline
 mixed algo. from $\widehat{\vb}_{pc}$ & $6.8$ s & $4.7$ s \\ 
 \hline
 mixed algo., random init. & $10.6$ s & $12.8$ s \\  
\hline
\end{tabular}
 \caption{\label{tab:t1} execution time for the setups with 4 measurement types.}
\vspace{-10mm}%Put here to reduce too much white space after your table 
\end{table}
\end{center}

\begin{center}
%\vspace{-3mm}
\begin{table}[H]
\begin{tabular}{c c c } 
\hline
   &$N_c=5000$ &$N_c=500000$ \\ [0.25ex] 
 \hline
 recursive algorithm & $0.17$ s & $0.17$ s  \\ 
 \hline
 $\mathscr{L}_{(\xb', \yb')}^{exact}$ min. from $\widehat{\vb}_{rec}$ & $44.4$ s & $10.9$ s \\ 
 \hline
 $\mathscr{L}_{(\xb', \yb')}^{exact}$ min., random init. & $272$ s & $94.4$ s \\ 
 \hline
 $\mathscr{L}_{(\xb', \yb')}^{gauss}$ min. from $\widehat{\vb}_{rec}$ & $38.8$ s & $16.7$ s \\ 
 \hline
 $\mathscr{L}_{(\xb', \yb')}^{gauss}$ min., random init. & $84.9$ s & $126.7$ s \\ 
 \hline
 mixed algo. from $\widehat{\vb}_{rec}$ & $47.8$ s & $26.1$ s \\ 
 \hline
 mixed algo., random init. & $62$ s & $38.4$ s \\  
\hline
\end{tabular}
 \caption{\label{tab:t1} execution time for the setups with 15 measurement types.}
\vspace{-7mm}%Put here to reduce too much white space after your table 
\end{table}
\end{center}

In conclusion, we recommend using the mixed algorithm for the likelihood, it is a good compromise between the  $\mathscr{L}_{(\xb', \yb')}^{gauss}$ minimization and the  $\mathscr{L}_{(\xb', \yb')}^{exact}$ minimization. The choice between the setup with $4$ types of measurements and the setup with $2n_{qb}+1$ types of measurements is less obvious. The first one is obviously simpler for the operator and the likelihood optimization is faster (see Table 1 and Table 2) but:
\begin{itemize}[leftmargin=3.5mm]
\item It yields a slightly less precise result. The median error with the mixed algorithm and $N_c=5000$ is $0.22$ against $0.19$ with 15 measurement types.
\item We have no closed-form algorithm that retrieves the state from the measurements. We must rely on PhaseCut which is unprecise. PhaseCut is also slow but the time gained during the mixed ML algorithm more than makes up for it (see Table 1 and Table 2).
\item We explained (in Section \ref{section:injectivity}) why we think the measurements are injective, and in practice all 1000 tested states were recovered, but we were unable to prove the injectivity so far.
\end{itemize}
\section{Conclusion and future work}
In this paper we first showed how some of the work made in the applied mathematics community in the field of phase recovery can be used to define a set of four types of $d$-outcome measurements that should be enough to achieve QST for any pure state using the PhaseCut optimization algorithm. We also proposed a set of $(2n_{qb}+1)$ types of $d$-outcome measurements as well as a recursive algorithm which allows explicit reconstruction of the state ($n_{qb}$ is the number of qubits, $d=2^{n_{qb}}$). Experimentally, they both give similar performances when the total number of measurements is the same (slight advantage for the second set of measurements); the first set is easier to set up and the second set is more theoretically sound.

The initial estimates of the considered state are then fined tuned with the maximum likelihood approach that is widely used in the quantum information processing literature. We introduced some refinements which make it more robust by considering a smooth an easy way to maximize an approximation of the likelihood.

We intend to use those QST methods to perform quantum process tomography (QPT) like in \cite{SSP}. In \cite{SSP} we introduced a QPT method that relies on measuring the state of the system after different time delays. At each time delay, we have to perform QST.

\appendix
\section{Covariance matrix and likelihood of the error on the sample probabilities}
\label{section:cov}

\subsection{Covariance matrix}
\label{section:cov1}
Appendix A aims at computing the asymptotic law of $\sqrt{N} \sbm{\varepsilon} =  \sqrt{N} (\widehat{\pb}-\pb)$ defined in Section \ref{section:gauss_ml} and at simplifying the expression of the likelihood of $\sbm{\varepsilon}$. We consider that $\pb$ contains the probabilities of a single type of $d$-outcome measurement. The generalization is straightforward as the errors on different measurements are independent (see Section \ref{section:extension}). The only random vector in $\sbm{\varepsilon}$ is $\widehat{\pb}$ defined as the vector that contains the sample probabilities of each of the $d$ outcomes. So $\widehat{\pb} = \frac{1}{N} \nb$ where each component $n_i$ of $\nb$ contains the number of times the $i$-th outcome occurred. By definition $\nb$ follows a multinomial distribution characterized by the number of trials $N$ and the theoretical probabilities of each outcome contained in $\pb$. The expected value and covariance matrix of the multinomial distribution are known: $E(\nb) = N\pb$ and $Cov(\nb) = N(diag(\pb)-\pb\pb^T)$.  

We want to use the central limit theorem so let us write $\nb$ as a sum: $\nb=\sum_{k=1}^{N}\deltab_k$ where the $\{\deltab_k\}_k$ are independent and have the same distribution for different $k$. $\deltab_k$ contains $d-1$ zeros and one $1$ at a random index $i_k\in \{1, ..., N\}$ whose density function is $j \longrightarrow p_j$ (i.e. the probability that $i_k$ takes the value $j \in \{1, ..., N\}$ is $p_j$, the $j$-th element of  $\pb$). $\deltab_k$ follows a multinomial distribution with $N=1$ trial. Its expected value is therefore $\pb$ and its covariance matrix is $diag(\pb)-\pb\pb^T$.
Therefore $\sbm{\varepsilon}$ is the difference between the empirical average of  $\deltab_k$ with $N$ realizations and its expected value. According to the central limit theorem, when $N\to +\infty$, the distribution of  $\sqrt{N} \sbm{\varepsilon}$ tends to a centered multivariate normal distribution, and its covariance matrix is $\sbv{\sbv{\Sigma}_{full}} = diag(\pb)-\pb\pb^T$. $\widehat{\sbv{\sbv{\Sigma}_{full}}}$ is an estimate of $\sbv{\sbv{\Sigma}_{full}}$, it uses $\widehat{\pb}$ as we do not want to depend on the unknown vector $\pb$: $\widehat{\sbv{\sbv{\Sigma}_{full}}} = diag(\widehat{\pb})-\widehat{\pb}\widehat{\pb}^T$. 
\subsection{Likelihood}
\label{section:cov2}
The easiest way to compute the likelihood of a vector that follows a multivariate normal distribution requires us to invert the covariance matrix \cite{cours_proba}. If the covariance matrix is not invertible, then it is not of full rank, this means that at least one component of the random vector is linearly dependent on the others and therefore it is not needed to compute the likelihood. Those components can be removed and the likelihood of the smaller vector is the same as the likelihood of the original vector. In our case, the components of $\sqrt{N} \sbm{\varepsilon}$ sum to zero, therefore its covariance matrix is not invertible and any component can be removed without loosing any information that could be used to compute the likelihood. Let us consider $\sqrt{N} \barbelow{\sbm{\varepsilon}}$, it is the same vector as $\sqrt{N} \sbm{\varepsilon}$ with the last component removed, and thus, its covariance matrix is the same with the last row and column removed: $\sbv{\Sigma} = diag(\barbelow{\pb})-\barbelow{\pb}\barbelow{\pb}^T$ ($\barbelow{\pb}$ is $\pb$ with the last element removed). It can be estimated with the sample probabilities $\barbelow{\widehat{\pb}} = \begin{pmatrix} \widehat{p_1} \\ \vdots \\ \widehat{p_{d-1}} \end{pmatrix}$ instead of $\barbelow{\pb}$. The resulting matrix is $\widehat{\sbv{\Sigma}} = diag(\barbelow{\widehat{\pb}})-\barbelow{\widehat{\pb}}\barbelow{\widehat{\pb}}^T$. Straightforward calculations show that if no element of $\widehat{\pb}=\begin{pmatrix} \widehat{p_1} \\ \vdots \\ \widehat{p_d} \end{pmatrix}$ (with $\widehat{p_d} = 1-\sum_{k=1}^{d-1} \widehat{p_k}$) is zero, then, $\widehat{\sbv{\Sigma}}$ is invertible and 

\begin{equation}
\label{eqn:sigma_hat}
\widehat{\sbv{\Sigma}}^{-1} = \frac{1}{\widehat{p_d}}\sbv{\mathds{1}}+diag(1/ \barbelow{\widehat{\pb}}).
\end{equation}

\noindent is its inverse. $1/ \barbelow{\widehat{\pb}}$ is the element-wise inverse of $\barbelow{\widehat{\pb}}$ and $\sbv{\mathds{1}}$ is the $d-1\times d-1$ matrix with only ones. In practice, elements of $\widehat{\pb}$ can be zeros, it would make the matrix singular. In order to overcome this difficulty and avoid giving too much importance to the errors on the scarcely observed outcomes, we modify the sample probability and create a new vector $\widetilde{\pb}$:
\begin{equation}
\widetilde{\pb} = \frac{\widehat{\pb}+\frac{5}{N}}{1+\frac{5d}{N}}. 
\end{equation}
This means that we consider that each outcome has been observed 5 more times than it actually was, and the total number of observations changes from $N$ to $N+5d$ (the choice of 5 is arbitrary). This is a standard method to make a criterion smoother (see \cite{edgedML}). The resulting estimate of the inverse of the covariance matrix is 
\begin{equation}
\label{eqn:sigma_tilde}
\widetilde{\sbv{\Sigma}}^{-1} = \frac{1}{\widetilde{p_d}}\mathds{1}+diag(1/ \barbelow{\widetilde{\pb}})
\end{equation}
With the inverse of $\widetilde{\sbv{\Sigma}}$ and knowing that the distribution is normal and centered, we can compute the negative log-likelihood of the vector (see \cite{cours_proba}):
\begin{equation}
\mathscr{L}_{(\xb', \yb')}^{gauss}(\widehat{\pb})= N \barbelow{\sbm{\varepsilon}} (\widehat{\pb}, \xb', \yb')^T \widetilde{\sbv{\Sigma}}^{-1} \barbelow{\sbm{\varepsilon}} (\widehat{\pb}, \xb', \yb').
\end{equation}
\noindent We use $\widehat{\pb}$ and not  $\widetilde{\pb}$ to compute $\barbelow{\sbm{\varepsilon}}$ otherwise estimator that minimizes the criterion would become biased (as the minimum of $\mathscr{L}_{(\xb', \yb')}^{gauss}$ would fit $\widetilde{\pb}$ which does not contains the actual sample probabilities) and the criterion would not be smoother.

Let us simplify this expression using (\ref{eqn:sigma_tilde}) and the fact that $\sum_k \varepsilon_k = 0 \Rightarrow \varepsilon_d=-\sum_{k=1}^{d-1} \varepsilon_k$:

$\begin{matrix} N \barbelow{\sbm{\varepsilon}}^T \widetilde{\sbv{\Sigma}}^{-1} \barbelow{\sbm{\varepsilon}} & = & N \barbelow{\sbm{\varepsilon}}^T \begin{pmatrix} \frac{1}{\widetilde{p}_d} \sum_{k=1}^{d-1} \varepsilon_k + \frac{\varepsilon_1}{\widetilde{p}_1}  \\ \vdots \\ \frac{1}{\widetilde{p}_d} \sum_{k=1}^{d-1} \varepsilon_k + \frac{\varepsilon_{d-1}}{\widetilde{p}_{d-1}} \end{pmatrix} \\ & = & N \barbelow{\sbm{\varepsilon}}^T \begin{pmatrix} \frac{\varepsilon_1}{\widetilde{p}_1} - \frac{\varepsilon_d}{\widetilde{p}_d}  \\ \vdots \\ \frac{\varepsilon_{d-1}}{\widetilde{p}_{d-1}}  - \frac{\varepsilon_d}{\widetilde{p}_d}  \end{pmatrix} \\ & = & N\big( \sum_{k=1}^{d-1} \frac{\varepsilon_k^2}{\widetilde{p}_k} - \frac{\varepsilon_d}{\widetilde{p}_d}  \sum_{k=1}^{d-1} \varepsilon_k\big) \\ & = & N \sum_{k=1}^{d} \frac{\varepsilon_k^2}{\widetilde{p}_k} .\end{matrix}$

\noindent Therefore, the expression of the negative log-likelihood is:
\begin{equation}
\mathscr{L}_{(\xb', \yb')}^{gauss}(\widehat{\pb})= N \sum_{k=1}^{d} \frac{\varepsilon_k (\widehat{\pb}, \xb', \yb')^2}{\widetilde{p}_k}.
\end{equation}
%This is the expression of the statistics of Pearson's chi squared test \cite{Pearson} (with $\widetilde{p}_k$ replaced by $p_k$ as there is no regularization in the Pearson's test). The main difference with what we do here is that we want to find the theoretical probability that maximizes the likelihood whereas the Pearson's test assumes that the theoretical probability is known and fixed and aims to test whether the difference between the two probabilities can arise by chance and to give a p-value. According to \cite{Pearson} the statistic of the test asymptotically approaches a $\chi^2$ with $d-1$ degrees of freedom. This is coherent with what we have because our version of the normalized log-likelihood of a Gaussian vector with $d-1$ component (like $\barbelow{\sbm{\varepsilon}}$) follows a $\chi^2$ with $d-1$ degrees of freedom. 
\subsection{Extension to several $d$-outcome measurements}
\label{section:extension}
Since the beginning of the appendix we assumed that only one type of measurement with $d$ outcomes was performed. In practice the methods we describe require either $4$ (in Section \ref{section:petitA}) or $2n_{qb}+1$ (in Section \ref{section:grandA}) types of measurements. The errors between the empirical and theoretical probabilities of different measurements are independent. Therefore if $\barbelow{\sbm{\varepsilon}} (\widehat{\pb}, \xb', \yb')$ contains $n_t>1$ types of measurements and $d n_t$ real components, then, its covariance matrix is a block diagonal matrix with the covariance matrix of each measurement type on the diagonal (because the measurement errors on two different measurement types are independent.). And the same goes for the inverse of its regularized covariance matrix:
\begin{equation}
\widetilde{\sbv{\Sigma}}^{-1} = \begin{bmatrix}  \widetilde{\sbv{\Sigma}}^{-1}_1 & & \\ & \ddots& \\ & & \widetilde{\sbv{\Sigma}}^{-1}_{n_t}\end{bmatrix}.
\end{equation}
Each $\widetilde{\sbv{\Sigma}}^{-1}_{k}$ is the regularized inverse of the covariance matrix for one measurement type defined in (\ref{eqn:sigma_tilde}).

The negative log-likelihood of $\sbm{\varepsilon} (\widehat{\pb}, \xb', \yb')$ containing $n_{prob} = n_t d$ measurements errors on $n_t$ types of measurements is the sum of the $n_t$ negative log-likelihoods of the error vectors of each measurement type
\begin{equation}
\mathscr{L}_{(\xb', \yb')}^{gauss}(\widehat{\pb})= N \sum_{k=1}^{n_{prob}} \frac{\varepsilon_k (\widehat{\pb}, \xb', \yb')^2}{\widetilde{p}_k}.
\end{equation}

\bibliography{mybibfile}

\end{document}